\documentclass[12pt,preprint,dvips]{aastex}
\usepackage[dvips]{color}

\def\ltsima{$\;\buildrel < \over \sim \;$}
\def\simlt{\lower.5ex \hbox{\ltsima}}
\def\gtsima{$\;\buildrel > \over \sim \;$}
\def\simgt{\lower.5ex \hbox{\gtsima}}
\newcommand{\asec}{$^{\prime\prime}$}
\newcommand{\amin}{$^{\prime}$}

\newcommand{\re}{\textcolor[rgb]{0,0,0}}

\newcommand{\etal}{{\em et$\:$al.}}

\shorttitle{Spitzer spectral line mapping of SNR}
\shortauthors{Neufeld et al.}

\begin{document}

\title{Spitzer spectral line mapping of supernova remnants:\\
 I.\ Basic data and principal component analysis}
\author{David A. Neufeld\altaffilmark{1}, 
David~J.~Hollenbach\altaffilmark{2}, Michael~J.~Kaufman\altaffilmark{3},
Ronald~L.~ Snell\altaffilmark{4}, Gary~J.~Melnick\altaffilmark{5},
Edwin~A.~Bergin\altaffilmark{6}, and Paule~Sonnentrucker\altaffilmark{1}}

\altaffiltext{1}{Department of Physics and Astronomy, Johns Hopkins University,
3400 North Charles Street, Baltimore, MD 21218}
\altaffiltext{2}{NASA Ames Research Center, Moffett Field, CA 94035}
\altaffiltext{3}{Department of Physics, San Jose State University, 1 Washington Square, San Jose, CA 95192}
\altaffiltext{4}{Department of Astronomy, University of Massachusetts, 710 North Pleasant Street, Amherst, MA 01003}
\altaffiltext{5}{Harvard-Smithsonian Center for Astrophysics, 60 Garden Street, 
Cambridge, MA 02138}
\altaffiltext{6}{Department of Astronomy, University of Michigan, 825 Dennison Building, Ann Arbor, MI 48109}

\begin{abstract}

We report the results of spectroscopic mapping observations carried out toward \re{small ($1^\prime \times 1^\prime$)} regions within the supernova remnants W44, W28, IC443, and 3C391 using the Infrared Spectrograph of the {\it Spitzer Space Telescope}.  These observations, covering the $5.2 - 37\rm\,\mu m$ spectral region, have led to the detection of a total of
15 fine structure transitions of Ne$^+$, Ne$^{++}$, Si$^+$, P$^+$, S, S$^{++}$, Cl$^+$, Fe$^+$, and Fe$^{++}$;
the S(0) -- S(7) pure rotational lines of molecular hydrogen; and the R(3) and R(4) transitions of hydrogen deuteride.   In addition to these 25 spectral lines,  
the 6.2, 7.7, 8.6, 11.3 \re{and 12.6} $\mu$m PAH emission bands were also observed.  Most of the detected line
transitions have proven strong enough to map in several sources, providing a comprehensive picture of the relative distribution of the various line emissions observable in the {\it Spitzer}/IRS bandpass.  A principal component analysis of the spectral line maps  reveals that the observed emission lines fall into five distinct groups, each of which may exhibit a distinct spatial distribution: (1) lines of S and H$_2\,(J > 2)$; (2) the H$_2$ S(0) line; (3) lines of ions with appearance potentials less than 13.6~eV; (4) lines of ions with appearance potentials greater than 13.6~eV, not including S$^{++}$; (5) lines of S$^{++}$.  Lines of group (1) likely originate in molecular material subject to a slow, nondissociative shock that is driven by the overpressure within the supernova remnant, and lines in groups (3) -- (5) are associated primarily with dissociative shock fronts with a range of (larger) shock velocities.  The H$_2$ S(0) line shows 
%both a shock-excited component and 
\re{a low-density diffuse emission component, and -- in some sources -- a shock-excited component.}

\end{abstract}


\section{Introduction}

Supernovae have long been recognized as a critical source of energy input to the ISM (e.g.\ Cox \& Smith 1974; McKee \& Ostriker 1977).  They send shock waves propagating through the ISM, creating large cavities filled with hot ionized material.  Eventually, supernova-driven shock waves become radiative, emitting strong line emissions -- primarily at optical and ultraviolet wavelengths -- that have been widely observed from supernova remnants (e.g.\ Weiler \& Sramek 1988).  Where supernova remnants encounter molecular clouds, they drive slower shock waves that compress and heat the molecular material and result in strong infrared line emission.  Depending upon the shock velocity, such shock waves may be either dissociative -- resulting in the destruction of molecules by collisional processes -- or nondissociative, in that molecules are heated but nevertheless survive the passage of the shock front.  The theory of interstellar shock waves, both in atomic and molecular media, has been the subject of extensive investigation (reviewed, for example, by Draine \& McKee 1993).

Using the Infrared Spectrograph (IRS) of the {\it Spitzer Space Telescope}, we have carried out infrared spectroscopic observations of four regions in which a supernova remnant interacts with a molecular cloud: W44, W28, 3C391, and IC443. 
The IRS covers the 5.2 -- 37$\,\rm \mu m$ region with unprecedented sensitivity, providing access to the lowest eight pure rotational transitions of molecular hydrogen together with fine structure lines of \re{[SI] and a variety of} atomic ions.  In the observations reported here, IRS spectroscopy indicated the presence of interstellar gas over a huge temperature range, from $\simlt 100$~K (H$_2$ S(0) $28\rm \,\mu m$ line) to $\simgt 10^5\, \rm K$ ([NeIII] 15.6 and $36\rm \,\mu m$ lines).

In this paper, we present a general overview of the data set we have obtained toward these four supernova remnants.  In \S 2, below, we give a brief overview of the four supernova remnants selected for this study.
In \S 3, we discuss the observations and methods used to reduce the observational data, and we present spectral line maps for a representative sample of the transitions that we detected.  In \S 4, we describe how Principal Component Analysis can be used to identify distinct groups of transitions that share a common origin in a particular component of the interstellar gas.  In \S 5, we discuss the similarities and differences in the spatial distributions of the various spectral lines we observed.   More detailed analyses of the different shocked gas components will follow in future papers.

\section{Source selection}

The four supernova remnants we have observed \re{-- W44, W28, 3C391 and IC443} -- represent 
classic examples of the interaction between supernova-driven shock waves and molecular clouds.  
Brief descriptions of each source are given below, and finder charts appear in Figures 1 -- 4.  
In each source, we have centered the mapped region at a position that is known from previous observations
to be a luminous source of molecular emissions; these are all positions that have been observed previously by 
\re{the Short- and/or Long Wavelength Spectrometer} of the {\it Infrared Space Observatory} ({\it ISO}).  \re{All four sources lie close to the Galactic plane, with the mapped regions lying at Galactic latitudes b$ \,\, \sim -0.43^\circ, -0.30^\circ, 0.03^\circ$ and 3.02$^\circ$ respectively for W44, W28, 3C391 and IC443 (see Table 1). }

\subsection{W44}

W44 lies at a distance of 2.5 kpc and has a radius of 11 -- 13 pc (Cox et al.\ 1999).  
The remnant is associated with the pulsar B1853+01, which is believed to be the progenitor 
star and provides an age of 20,000 years (Wolszczan et al.\ 1991; based upon the ratio of the pulsar period to the period derivative). W44 is a prototype for a 
new class of supernova remnants called ``mixed-morphology" SNRs, because the radio continuum images 
exhibit a shell-like morphology while the center is filled in with X-ray emission 
(Jones et al.\ 1993; Rho \& Petre 1998).   Such morphologies have been attributed to the 
interaction between the supernova shock front and a molecular cloud (Chevalier 1999; 
Yusef-Zadeh et al.\ 2003), and W44 is indeed associated with a Giant Molecular Cloud, recently 
mapped in CO emission (Reach, Rho, \& Jarrett 2005: hereafter RRJ05).  

\re{RRJ05} and Cox et al.\ (1999) provide nice summaries of the available observations.  
In general, there is strong evidence for supernova interaction with a dense molecular cloud  
through the detection of OH 1720 MHz maser emission, broad (FWHM $\sim$ 20 -- 30 km/s) CO molecular 
line emission, vibrational and rotational lines of molecular hydrogen, and atomic fine structure lines (Claussen et al.\ 1997; Seta et al.\ 1998; Reach \& Rho 2000;  RRJ05).    Comparison to shock models are consistent with a mixture of dissociative and non-dissociative shocks impacting on gas with densities of $100-10^4 \, \rm  cm^{-3}$ (Reach \& Rho 2000; RRJ05), although Cox et al.\ (1999) present a model that matches most data but requires no dense cloud interaction.    The region we have observed with {\it Spitzer} (Figure 1) lies at the northeast edge of the W44 supernova remnant, coincident with OH maser Region E (Claussen et al.\ 1997).
Observations of CO and X-rays imply total hydrogen column densities of $2 \times 10^{22}\, \rm cm^{-2}$  along the sight-line (Rho et al.\ 1994; RRJ05), corresponding to a visual extinction $A_V \sim 10$~mag.

\subsection{W28}

The supernova remnant W28 is located in a complex region of the 
Galactic disk where a number of nearby large \ion{H}{2} regions 
(M8 and M20) and young clusters (e.g., NGC 6530) are present (Goudis 1976). 
Kinematic considerations based on molecular line broadening place the remnant 
at a distance of 1.9 $\pm$ 0.3 kpc (Vel\'azquez et al. 2002). The distribution 
of radio molecular emission lines indicates that the remnant has an irregular 
shell-like structure with prominent peak emissions in the northern and northeast 
sides of the shell where an interaction between the remnant and a nearby molecular 
cloud was suggested (Wootten 1981). Further evidence for such an interaction arises 
from the detection of numerous discrete OH 1720 MHz maser spots along the northeast 
ridge where the remnant is believed to be interacting with a molecular cloud 
(Claussen et al. 1997). This maser activity has been shown to constitute a tracer of 
supernova remnant-molecular cloud interaction (Yusef-Zadeh et al. 2003). Detection of 
the H$_2$ S(3) and S(9) lines as well as a number of ionic species with {\it ISO} suggests 
the presence of shocks in the northern part of the remnant (Reach \& Rho 2000) as well. 
In the optical, short emission filaments are observed superposed over a more diffuse and 
centrally-peaked emission seen throughout most of the interior of W28 (van den Bergh 1973; 
Long et al. 1991). Diffuse thermal X-ray emission is also seen throughout the region 
delineated by the radio shells and shows a rather patchy but centrally-peaked emission 
distribution (Dubner et al. 2000; Rho \& Borkowski 2002). Like W44, W28 shows 
a ''mixed-morphology" that combines a shell-like radio structure with a 
centrally-filled diffuse X-ray emission structure. Observations of optical [SII]
line ratios (Long et al.\ 1991) suggest a reddening of $E(B-V) \sim 1 - 1.3$~mag, 
corresponding to a visual extinction of $3 - 4$~mag for an
$A_V / E(B-V)$  ratio of 3.1.

\subsection{3C391}

3C391 (G31.9+0.0) is an X-ray- and radio-bright supernova remnant that 
lies in the Galactic plane.  The X-ray emission is diffuse (i.e., no limb
brightening) and centered within the radio shell (Rho \& Petre 1996).  
$^{12}$CO $J\,=\,$1$-$0 maps toward 3C391 reveal the presence of moderately
dense ($n$(H$_2$)$\,\sim\,$300 -- 1000~cm$^{-3}$) molecular gas along the northwest
portion of the remnant (Wilner, Reynolds \& Moffett 1998).  The radio 
continuum emission from the remnant (Reynolds \& Moffett 1993), shown in Fig.~3
with the region mapped by {\it Spitzer} superposed, suggests evolution
along a strong density gradient, with compression of the supernova blast wave to 
the northwest in the direction of the strongest CO emission and lower continuum
surface brightness emission to the southeast away from the molecular gas. 
It is likely that the progenitor star exploded within a dense molecular
cloud and the expanding gas has now broken out of the cloud to the 
southeast.

Evidence exists for both dissociative and nondissociative shocks toward
3C391 (Reach et al.\ 2002). Narrowband filter images in the [Fe$\:$II] 
1.64$\:\mu$m line show a pronounced intensity peak along the bright 
radio bar (18$^{\rm h}$ 49$^{\rm m}$ 16$^{\rm s}$, --0$^{\rm o}$ 55\amin\
00\asec) at the interface between the supernova remnant and the molecular cloud.
Similarly, the ISOCAM CVF 5 -- 18$\:\mu$m spectrum of the same region is dominated by
emission lines from atomic ions -- such as [Fe$\:$II] 5.5$\:\mu$m,
[Ar$\:$II] 6.9$\:\mu$m, [Ne$\:$II] 12.8$\:\mu$m, and [Ne$\:$III] 15.5$\:\mu$m --
while the mid-infrared H$_2$ rotational lines in the same spectrum
are either weak or absent (Reach et al.\ 2002).  Approximately 2.8\amin\ to the south and east of 
this position, near
$\alpha=$18$^{\rm h}$ 49$^{\rm m}$ 22$^{\rm s}$, $\delta=-0^{\rm o}$ 57\amin\ 22\asec\ (J2000),
continuum-subtracted H$_2$ 2.12$\:\mu$m images show an $\sim\,$30\asec\ diameter
area containing a cluster of small ($<\,5$\asec) clumps of H$_2$ emission. 
The ISOCAM CVF spectrum of this region
shows strong H$_2$ S(2) through S(7) H$_2$ line emission (Reach et al.\ 2002).
Longer wavelength {\it ISO} LWS spectra toward this location show thermal H$_2$O, 
OH, high-$J$ $^{12}$CO (Reach \& Rho 1998) and [O$\,$I] 63$\:\mu$m (Reach \& Rho 1996)
emission.  This so-called broad molecular line region (see Reach \& Rho 1999)
is also the site of OH 1720~MHz maser (Frail et al.\ 1996) and
broad (FWHM $\sim\,$20 km~s$^{-1}$) $^{12}$CO ($J=2-1$), CS ($J=3-2$), CS ($J=5-4$),
and HCO$^+$ ($J=1-0$) emission (Reach \& Rho 1999).  The existence of strong and
broad molecular line emission has led to the conjecture \re{that the supernova 
blast wave} is encountering dense ($n$(H$_2$)$\,\sim\,$10$^4\,$--$\,$10$^5$~cm$^{-3}$)
cores present within the molecular cloud from which the progenitor of 3C391
formed.  The region that we have mapped with {\it Spitzer} (Figure 3) 
is centered on the broad molecular line region.

The extinction to 3C391 is very high due to its location in the Galactic
plane and its 9~kpc distance (Caswell et al.\ 1971; Radhakrishnan et al.\ 1972).  
Column densities
inferred from the spectral analysis of ROSAT data yield a foreground column
density of 2 -- 3.6 $\times$ 10$^{22}$ cm$^{-2}$ (Rho \& Petre 1996).  Since this
analysis has been applied by averaging spectra obtained over relatively large
areas of the remnant, it is likely that the higher value applies to the denser
broad molecular line region of interest here.  Thus, assuming a foreground
column density of 3.6 $\times$ 10$^{22}$ cm$^{-2}$, a line-of-sight visual
extinction of $A_{\rm v}\:=\:$19 magnitudes is derived (Reach et al.\
2002).

\subsection{IC443}

IC443 is a well-studied SNR located in the outer 
Galaxy at a distance of approximately 1.5 kpc 
(Georgelin 1975; Fesen \& Kirshner 1980; Fesen 1984).
It is one of the more
striking examples of the interaction of a SNR with
the interstellar molecular medium.  IC443C, the position
that we have targeted with {\it Spitzer}, is one of 
several shocked gas regions within the SNR first identified 
by DeNoyer (1978) based on the presence of high velocity 
HI emission.  These shocked gas regions were later
detected to have broad OH absorption and CO emission 
lines (DeNoyer 1979a,b).  More extensive 
observations of CO and HCO$^+$ emission in IC443 
by Dickman et al.\ (1992) revealed many more regions
of broad molecular line emission and showed
that they formed an expanding, tilted ring. This ring
of shocked gas presumably delineates where the SNR shock 
has encountered the molecular 
interstellar material.  This ring is also seen 
in 2 $\mu$m H$_2$ emission (Burton et al.\ 1988; Richter et al.\ 1995b).

\re{Based upon the kinematics of the observed CO and HCO$^+$ emissions}, IC443C is believed to be located 
on the front side of the expanding ring (Dickman et al.\ 1992).  
The unshocked molecular gas in this direction 
has a v$_{LSR}$ of --3 km s$^{-1}$.  The broad velocity
emission from the shocked molecular gas extends from the
velocity of the unshocked gas to a v$_{LSR}$ of 
--70 km s$^{-1}$ (Snell et al.\ 2005).  The region of most intense molecular
broad-line emission is about 1 arcminute in size.

Emission from the shocked interstellar gas in IC443 
has been extensively studied and detected in numerous atomic
and molecular tracers, including HI (Braun \& Strom 1986) 
vibrationally excited H$_2$ (Burton et al.\ 1988), the 63 $\mu$m
line of [OI] (Burton et al.\ 1990), the pure rotational lines
of H$_2$ (Richter et al.\ 1995a; Cesarsky et al.\ 1999) and numerous molecular
species (DeNoyer \& Frerking 1981; Ziurys et al.\ 1989; van Dishoeck et al.\ 1993).
Most recently, the detection of water was reported in several of
these shocked regions, including IC443C (Snell et al.\ 2005).  
Most studies have concluded that a combination of 
shocks are needed to explain the emission seen in these 
atomic and molecular tracers.  Snell et al.\ (2005) suggested 
the presence in IC443C of both a fast J-shock and either a 
slow C-shock or slow J-shock.  Given the observed ratio of H$_2$ vibrational 
lines, Richter et al.\ (1995b)
derived 2.1 $\mu$m extinctions of 1.3 and 1.6 mag for two slit positions
in IC443C, corresponding to visual extinctions $A_V$ of 12 and 15 mag.

\section{Observations and results}

W44, W28, 3C391, and IC443C were each observed using the IRS as part of a Cycle 2 General Observer Program, the Short-Low (SL), Short-High (SH) and Long-High (LH) modules being used to obtain complete spectral coverage over the wavelength range available to IRS.  \re{The spectral resolving power of IRS, $R = \lambda / \Delta \lambda$, ranges from 60 to 127 for the SL module and is $\sim 600$ for the SH and LH modules.}  Spectral line maps of size $\sim 1^\prime \times 1^\prime$ were obtained by stepping the slit perpendicular \re{to its long axis by one-half its width}, and -- in the case of SH and LH -- parallel to its long axis 
\re{by 4/5 of its length}.  Table 1 summarizes the observational details for each source; 
\re{note that the map center positions are not in all cases identical to the (0,0) positions defined in the captions to Figures 10 -- 13.}

The data were processed at the Spitzer Science Center (SSC), using version 13.2 of the processing pipeline, and then reduced further using the SMART software package (Higdon et al.\ 2004), supplemented by additional routines (Neufeld et al.\ 2006a), that provide for the removal of bad pixels in the LH and SH data, the calibration of fluxes obtained for extended sources, the extraction of individual spectra at each position sampled by the IRS slit, and the creation of spectral line maps from the set of extracted spectra.  \re{No background subtraction could be performed, since no off-source measurements were made;  while background subtraction might have affected the measured continuum fluxes, it would be expected to have no effect upon the derived intensities for {\it spectral lines}. There is no evidence for fringing in any of the spectra.}

For several transitions observed in the LH and SH modules, the resultant line maps exhibited striping perpendicular to the long axis of the slit.  We determined this effect to be an artifact created by the standard algorithm \re{used in the IRSCLEAN software} to obtain interpolated data for bad pixels; that algorithm interpolates in the dispersion direction, causing a systematic underestimate in the intensity when a bad pixel falls close to the central wavelength of a spectral line.  The striping was successfully removed by applying a correction factor, computed separately for each position along the slit and for each affected spectral line.  \re{The required correction factors were obtained by considering -- separately for each of the five positions along the slit -- the cumulative distribution functions for the line fluxes obtained over the entire map.  In other words, for each five possible positions along the slit, we plotted the fraction of line flux measurements that were smaller than a given value, as a function of that value.  Making the assumption that the cumulative distribution functions should be similar except for the highest flux values, we computed the necessary correction factors needed to bring the five distribution functions into agreement.}

In order to improve the signal-to-noise ratio beyond that obtained at a single position, we have computed average spectra for each of the sources we observed.  The resultant spectra, plotted in Figures 5 -- 8, are the averages of all spectra within a circular region, the contributing spectra being weighted by a Gaussian taper (HPBW of 25$^{\prime\prime}$) from the center of each synthesized beam.  The central positions are given in the figure captions.  

The spectra plotted in Figures 5 -- 8 reveal a total of 25 spectral lines, detectable in one or more source.  These include 15 fine structure transitions of the species Ne$^+$, Ne$^{++}$ (2 transitions), Si$^+$, P$^+$, S, S$^{++}$ (2 transitions), Cl$^+$, Fe$^+$ (5 transitions), and Fe$^{++}$; 8 transitions of H$_2$ (the S(0) through S(7) pure rotational transitions), and the R(3) and R(4) transitions of HD (detected toward IC443C and reported previously by Neufeld et al.\ 2006b).  Beam averaged intensities are listed in Table 2.  In addition to these emission lines, the 6.2, 7.7, 8.6, 11.3 \re{and 12.6} $\mu$m PAH emission bands are readily apparent in the spectra of W44, W28, and 3C391, and the 11.3 $\mu$m micron feature is detected in IC443C.  A discussion of the PAH emission features is deferred to a future paper.

The observed line transitions represent a wide variety of excitation conditions.  In Figure 9, we show their position in a scatter plot, where the horizontal axis shows the appearance potential -- equal to zero for neutral species and equal to the ionization potential of X$^{(n-1)+}$ for the case of ion X$^{n+}$ -- and the vertical axis shows the critical density at which the rate of collisional de-excitation equals the spontaneous radiative rate.  In computing the critical density, we assumed the dominant collision partner to be electrons for ionized species and hydrogen molecules for neutral species.
   
In Figures 10 -- 13, again for each source, we present maps of a representative selection of six strong spectral lines:  the H$_2$ S(0), [SI] $\rm 25\,\mu m$, H$_2$ S(2), [SiII] $\rm 35\, \mu m$, [NeII] $\rm 12.8\, \mu m$, and [SIII] $\rm 33\, \mu m$ lines.  
%Most of the other spectral lines listed in Table 2 are also strong enough to be mapped, and additional
%spectral line maps are available in the online materials for this paper.  
As demonstrated by the Principal Component Analysis described in \S 4 below, each of the other transitions shows a spatial variation \re{that is almost identical} to one of the transitions plotted in Figures 10 -- 13.  \re{For example, except for a constant scaling factor, the [FeII] $\rm 26\, \mu m$ line maps are visually indistinguishable from those shown here for [SiII] $\rm 35\, \mu m$.  For the maps obtained with the SL module, the effective integration time and resultant signal-to-noise ratio is the same everywhere in the map.  For the SH and LH modules, the effective integration time is constant, except in overlap regions between adjacent pointings where the integration time is larger by a factor two.} 

\section{Principal Component Analysis}

A cursory inspection of Figures 10 -- 13 reveals both similarities and differences in the spatial distributions of the various emission lines that we have observed.  The method of Principal Component Analysis (PCA) provides a valuable tool for describing those similarities and differences.   The value of PCA in analysing multitransition maps was first demonstrated by Ungerechts et al.\ (1997), who applied the method to molecular line maps of the Orion Molecular Cloud and gave a comprehensive description of PCA that will not be repeated here (see also Heyer \& Schloerb 1997).  To summarize, PCA creates an orthogonal basis set of maps -- known as principal components (PC) -- such that each observed map can be represented as a linear combination of the PC.  The key feature of the new basis set is that the PC are obtained in decreasing order of amplitude: the first PC contains as much of the spatial variation as possible, the second PC - constructed orthogonal to the first - contains as much of the remaining variation as possible, and so forth.

Following Ungerechts et al.\ (1997), we applied the Principal Component Analysis to spectral line maps that had been mean-subtracted and divided by the standard deviation, \re{the latter being dominated by structure in the source rather than instrumental noise}.   This procedure gives equal weighting to each line instead of allowing the brightest lines to dominate the analysis.  \re{Because this analysis involves data obtained with all three modules (i.e.\ SL, SH, and LH), it was applied only to the intersection of the regions mapped by the different modules.}  
Applying PCA to the four sources we have observed, we find that the first PC contains $58 - 73\%$ of the total spatial variation in the maps, the first and second PCs contain together $82 - 88\%$, and the first three PCs contain $90 - 93 \%$ of the information.   In Figures 14 -- 17 we present the results of the PCA for each source.  The upper panel shows the fraction of the variance accounted for by each component (asterisks), along with the cumulative fraction for the first $n$ components as a function of $n$ (open squares).   The lower panels show the coefficients for the 1st, 2nd and 3rd PC needed to approximate the maps of each transition; the coefficients for the 1st and 2nd PC are shown in the left lower panel and those for the 2nd and 3rd PC are shown in the right lower panel.  (These figures were referred to as h-plots by Ungerechts et al.\ 1997.)  Because the PCA is applied to mean-subtracted line maps, the coefficients may be either positive or negative.   \re{While the 3rd PC contains information about the source -- as evidenced by the fact that transitions with similar coefficients for the 2nd PC tend to show similar coefficients for the 3rd -- the 4th and higher PCs are likely dominated by noise.} 

The principal components are normalized such that quadrature sum of the coefficients for each transition is unity.  Thus, if a particular transition were completely accounted for by the 1st and 2nd PC, its position in the lower left plot would lie on the circle of unit radius; similarly, if a particular transition were completely accounted for by the 2nd and 3rd PC, its position in the lower right plot would lie on the circle of unit radius.  The proximity of most points in the lower left plot to the circle of unit radius is a reflection of the fact that the first two PCs account for $\sim 85 \%$ of the variance in the observed line maps.  In the limit where two points actually lie on the unit circle, the cosine of the angle between the two radii measures the degree of \re{linear} correlation; points lying next to each other on the unit circle would be perfectly correlated (cos$\,\theta=+1$), while those lying exactly on opposite sides of the circle would be perfectly anticorrelated (cos$\,\theta=-1$).  For example, in W44, the lower left panel indicates that the [FeII] 26$\rm \,\mu m$ and [SiII] transitions are very well correlated with each other, strongly anticorrelated with both observed [SIII] transitions, and almost uncorrelated with the H$_2$ S(0) transition.  

In W44, the PCA yields a clear separation into five groups of spectral lines: (1) lines of S, and H$_2\,(J > 2)$; (2) the H$_2$ S(0) line; (3) lines of ions with appearance potentials less than 13.6~eV (Fe$^+$, Si$^+$,
P$^+$); (4) lines of ions with appearance potentials greater than 13.6~eV (Ne$^+$, Ne$^{++}$, Fe$^{++}$), not including S$^{++}$; (5) lines of S$^{++}$.  In the other sources, some of these groups share a common distribution, but in no case do the spectral lines {\it within} a group defined above fail to be strongly correlated with each other.  \re{In W28, lines in groups (1) and (2) share a similar distribution, which is markedly different from that of groups (3) -- (5).  In 3C391, lines in groups (1), (2), (3) and (5) show distributions that are readily distinguishable from each other, but group (4) shows a distribution that is very similar to that of group (3).  In IC443, groups (1), (3), and (4) are readily distiguishable from each other; group (2) (the H$_2$ S(0) line) has a similar distribution to that of group (1); and group (5) ([SIII] transitions) is not detected.
These behaviors suggest} that the line emissions we have observed originate in at least five physically distinct components (some of which have similar distributions -- as projected onto the sky -- in sources other than W44).   The implications of these groupings are discussed further in \S 5 below.

\section{Discussion}

\subsection{Line emission from neutral species: H$_2$ and atomic sulphur}

Each of the sources we observed shows strong emission in the H$_2$ S(0) through S(7) lines.  
In Figure 18, we present rotational diagrams derived from the H$_2$ line fluxes listed in Table 1.  Here we
corrected for extinction, adopting the foreground extinction estimates given in \S 2 and the interstellar extinction curves of Weingartner \& Draine (2001; ``Milky Way, $R_V=3.1$").
The rotational diagrams bear a strong similarity to those obtained previously from {\it Spitzer} observations of the Herbig-Haro objects HH54 and HH7 (Neufeld et al.\ 2006a).  In each case, a positive curvature in the rotational diagrams indicates the presence of an admixture of gas temperatures, while a characteristic zigzag pattern -- more pronounced for low-lying transitions -- indicates a nonequilibrium ortho-to-para ratio (OPR), smaller\footnote{Except in IC443C, where -- given the uncertainties -- the data are consistent with an equilibrium OPR of 3.}
than 3.  In Neufeld et al.\ (2006a), we argued that the nonequilibrium H$_2$ ortho-to-para ratios were consistent with shock models in which the gas is warm for a time period shorter than that required for
reactive collisions between H and para-H$_2$ to establish an equilibrium OPR.

As in Neufeld et al.\ (2006a), we obtained a two-component fit to the rotational diagrams, in which a warm component with temperature, $T_w$, ortho-to-para ratio, OPR$_w$, and column density $N_w$ coexists with a hot component with temperature, $T_h$, ortho-to-para ratio, OPR$_h$, and column density $N_h$.  The resultant parameters, listed in Table 3, are broadly similar to those obtained for HH7 and HH54,
and are consistent with the presence of nondissociative shocks\footnote{Given the magnetic field strengths typical of the interstellar medium, such shocks are expected to be of ``continuous" or ``C-type" in the designation of Draine (1980), and are qualitatively different from the ``jump" or ``J-type" shock fronts associated with faster, dissociative shock waves.  C-type shocks are characterized by continuous variations in the gas velocity, large ion-neutral drift velocities, and lower peak gas temperatures than those attained behind J-type shocks of the same velocity.} of velocity 10 -- 20$\, \rm km \, s^{-1}$.    Our two-component fits are based upon a simple LTE (local thermodynamic equilibrium) treatment of the H$_2$ excitation, a treatment that is accurate for the lower-lying transitions but may be an oversimplification in the case of the highest transitions (S(6) and S(7)).  In a future paper, we will present a refined analysis, including non-LTE effects, together with maps of the various physical parameters derived from the H$_2$ (and, for IC443C, the HD) maps.

While the H$_2$ S(1) through S(7) lines show a clear morphological similarity to each other, a result that is apparent from inspection of the spectral line maps and is confirmed by the PCA, the H$_2$ S(0) line is typically more extended than -- and, in \re{W44} and 3C391, essentially uncorrelated with -- the other H$_2$ rotational lines.  The [SI] 25$\,\rm \mu m$ line, by contrast, is always well correlated with the H$_2$ S(1) through S(7) lines, despite the fact that its upper state energy is very similar to that of H$_2$ S(0).  These behaviors are demonstrated clearly in Figure 19, which present scatter diagrams showing correlations between [SI] and H$_2$ S(0).  Each point in Figure 19 shows the mean intensities within a $5^{\prime \prime} \times 5^{\prime \prime}$ square region. The results shown for W28 are very revealing: while [SI] and H$_2$ S(0) are well correlated in W28 \re{(with a linear correlation coefficient of 0.88)}, the best-fit linear regression yields a positive x-intercept, indicating that there is an extended component of H$_2$ S(0) emission (with typical intensity $\rm \sim 2.5 \times 10^{-5} erg \, cm^{-2} \, s^{-1} \, sr^{-1}$) that is unassociated with measurable [SI] emission.   Figure 19 \re{suggests an explanation for why} [SI] $25 \, \rm \mu m$ and H$_2$ S(0) are \re{reasonably} well-correlated in IC433C and W28 
but not in W44 and 3C391: in W28 and IC443C, the shock excited contribution dominates the variations in the extended component, whereas in W44 and 3C391 it does not.  
  
The much lower [SI]/H$_2$ S(0) line ratio in the extended component could indicate a much lower abundance of atomic sulfur, or - more likely - a lower gas density, the critical density for the [SI] $25 \, \rm \mu m$ line being much greater than that of the 
H$_2$ S(0) line (see Fig.\ 9).  The extended H$_2$ S(0) emission in these sources has an intensity comparable to that measured by Falgarone et al.\ (2005), who observed H$_2$ pure rotational emission along a long sight-line within the Galaxy that did not intersect any region of high-mass star formation.  Falgarone et al.\ argued that the observed H$_2$ emission originated in a warm gas component (unassociated with supernova remnants or regions of star formation) that is present within the cold, quiescent ISM.  Alternatively, or in addition, the H$_2$ S(0) emission within supernova remnants might originate within low density photodissociation regions that are irradiated and heated by UV radiation from nearby fast shocks.  In contrast to the continuum radiation incident upon typical PDRs, the UV radiation emitted by fast shocks is dominated by spectral lines.  Although detailed calculations will be needed to model PDRs that are irradiated by UV emission from fast shocks, we speculate that a line-dominated spectrum may be particularly effective in exciting H$_2$ rotational emissions since -- except for spectral lines that happen to coincide with an H$_2$ Lyman or Werner band transition -- line radiation can heat the gas without photodissociating molecules. 
  
It is also noteworthy that the [SI] emission is better correlated with the warm H$_2$ than with the ionized species in these sources.  Haas et al.\ (1991) had previously suggested that [SI] was a tracer of fast dissociative shocks, rather than the slower nondissociative shocks responsible for the H$_2$ emissions.  At least in the supernova remnants that we have mapped with {\it Spitzer}, the PCA strongly suggests that [SI] traces slow nondissociative shocks. 

The atomic sulfur abundance can be estimated from the [SI] $25 \, \rm \mu m$ line flux.  Unfortunately, the critical density for the [SI] $25 \, \rm \mu m$ line is probably large in neutral gas 
($\sim 10^6 \rm \, cm^{-3}$ according to Hollenbach \& McKee 1989, Table 5), so the derived SI abundance depends on both the gas density and the adopted rate coefficient for excitation of SI in collisions with H$_2$.
IC443C presents the best case for determining the atomic sulfur abundance, because a measurement of 
the HD R(4)/R(3) line allows the gas pressure to be determined for this case (Neufeld et al.\ 2006b).
Adopting the H$_2$ column densities and temperatures obtained from a two-component fit to the rotational diagram, assuming the warm and hot gas components to be at a common pressure adjusted to match the HD R(4)/R(3) line ratio (Neufeld et al.\ 2006b), and making
use of Hollenbach \& McKee's (1989; hereafter HM89) estimate of the rate coefficient for excitation of [SI] fine structure collisions\footnote{This estimate is actually for excitation by atomic hydrogen, but -- since we are not aware of any estimate in the literature for collisional excitation of SI by H$_2$ -- we adopt it for excitation by H$_2$ as well.}, we derive an atomic sulfur abundance of $\sim 10^{-6}$ relative to H$_2$.  This value corresponds to $\sim 4\%$ of the solar abundance of sulfur ($1.4 \times 10^{-5}$ relative to H nuclei; Asplund, Grevesse, \& Sauval 2005).  The uncertainty in this estimate is large, not least because of the absence of a reliable estimate for the excitation rate coefficient.
Owing to the presence of a significant UV radiation field produced by fast dissociative shocks in the vicinity of the warm molecular component, Snell et al.\ (2005) argued that sulfur could be significantly ionized within the nondissociative molecular shocks that are responsible for the pure rotational emissions from H$_2$; thus the observed SI abundance might be entirely consistent with a negligible depletion of sulfur onto grains and a negligible fraction of the sulfur nuclei being bound within gas-phase molecules. Further discussion of the atomic sulfur abundance, and its implications for the chemistry of interstellar sulfur molecules, is deferred to a future paper.

\subsection{Line emission from ionized species} 

The PCA shows a very tight correlation between the [FeII], [PII] and [SiII] line intensities
in all the regions we observed, a result confirmed by the correlation plots shown in Figures 20 ([FeII] versus [SiII] intensity) and 21 ([PII] versus [SiII] intensity).  These close correlations are 
unsurprising, given the similar appearance potentials (all $< 13.6$~eV) of Fe$^+$, P$^+$ and Si$^+$: all three ions are expected behind shocks that are fast enough to cause significant ionization, and
can also be produced in the ionizing ``precursors" of fast shocks, where ultraviolet radiation propagates upstream and can photoionize the gas before it even enters the shock front.  
For shocks propagating in molecular gas, the calculations of HM89 suggest that a shock velocity of at least $35\, \rm km s^{-1}$ is required to produce detectable fine structure emissions from [FeII] and [SiII].  

One additional feature of the data, however, which is not revealed by our PCA of mean-subtracted line maps, is that the correlation plots typically show positive intercepts on the horizontal [SiII] axis.  As in the case of H$_2$ S(0), this behavior implies the presence of an extended gas component within which [SiII] is
enhanced relative to [FeII] and [PII].
%intensity ratios are larger than those in the shocked gas component.  
%Once again, extended  emission from lower density gas provides a possible explanation, the critical density 
%for [SiII] being smaller than that for [FeII].  
    
Four ions of appearance potential greater than 13.6~eV are detected in one or more region: Ne$^+$, Ne$^{++}$, Fe$^{++}$ and S$^{++}$.  Such ions, \re{which have been observed widely in supernova remnants (e.g.\ Raymond et al.\ 1997), are produced only in faster shocks.  For example, when shocks propagate in molecular gas
clouds, velocities 
$\simgt 80 \, \rm km s^{-1}$ are needed to ionize Ne (HM89).}  
These lines are generally well-correlated with each other, with the curious exception of the two [SIII] lines (19 and 33 $\rm \mu m$); in W44, the [SIII] lines are {\it anticorrelated} with the other emissions from ionized species.  We speculate that this behavior could represent an excitation effect, the critical density for the [SIII] transitions being smaller than those for [NeII] and [NeIII].  The [SIII] 18.7/33.5~$\mu$m and [NeIII] 15.6/36.0~$\mu$m line ratios both provide useful constraints on the electron density in the emitting region (Alexander et al.\ 1999).  The observed [SIII] 18.7/33.5~$\mu$m line ratio was 0.78, 
0.54 and 0.55 respectively in W44, W28, and 3C391, the three sources in which [SIII] emissions were detected.  All three values are consistent with the low density limit obtained by Alexander et al.\ (1999;
0.6 at a temperature of 10$^4$~K, or 0.7 at a temperature of $2 \times 10^4$~K).
Conservatively estimating the likely error on the ratio as $\pm 35\%$, assuming that the gas temperature is at least 10$^4$~K, and making use of the results presented by Alexander et al.\ (1999; their Figure 3), we obtain upper limits on the electron density of 600, 200, and 200 $\rm cm^{-3}$ for W44, W28, and 3C391.  \re{For all three sources, our best estimate of the [SIII] 18.7/33.5~$\mu$m line ratio is somewhat smaller than -- but probably consistent with -- that obtained by Oliva et al.\ (1989) from {\it ISO} observations of the young supernova remnant RCW 103 ($0.89 \pm 0.18$).}
The observed [NeIII] 15.6/36.0~$\mu$m ratio was $\sim 11$ in both W28 and 3C391 (and was consistent with 11 in W44 and IC443C \re{where} the 36.0~$\mu$m transition was not detected).  Once again, this value \re{is similar to that observed by Oliva et al.\ (1989) toward RCW 103 ($10 \pm 2$), and} is consistent with the low density limit obtained by Alexander et al.\ (1999)
(11 at temperature of both 10$^4$~K and $2 \times 10^4$~K), but in this case the implied upper limit on the electron density is $\sim 10^4 \, \rm cm^{-3}$ for both W44 and 3C391.  Thus the observed 18.7/33.5~$\mu$m and [NeIII] 15.6/36.0~$\mu$m line ratios are entirely consistent with a picture in which the [SIII] emissions originate in a spatially distinct gas component of lower density than that responsible for the observed [NeIII] emissions.  
In a future paper, we will present a more detailed analysis of the emission from ionized species in these sources, in the context of theoretical models for fast shocks and PDRs.  

\section{Summary}

\noindent 1) We have carried out spectroscopic mapping observations, covering the $5.2 - 37\rm\,\mu m$ spectral region, toward the supernova remnants W44, W28, IC443C, and 3C391, with the use of the Infrared Spectrograph (IRS) \re{of} the {\it Spitzer Space Telescope}.  In each case, a region $\sim 1^\prime \times 1^\prime$ has been mapped.  \re{Except in 3C391, which is more distant and has a much smaller angular size than the other objects we observed, the mapped regions cover less than 1 percent of the total radio-emitting area of the supernova remnant.}

\noindent 2) These observations, performed using the Short-Low, Short-High and Long-High modules of the IRS, have led to the detection of a total of
15 fine structure transitions of Ne$^+$, Ne$^{++}$ (2 transitions), Si$^+$, P$^+$, S, S$^{++}$ (2 transitions), Cl$^+$, Fe$^+$ (5 transitions), and Fe$^{++}$;
the S(0) -- S(7) pure rotational lines of molecular hydrogen; and the R(3) and R(4) transitions of hydrogen deuteride (reported previously by Neufeld et al.\ 2006b).  

\noindent 3) We have performed a principal component analysis (PCA) of the spectral line maps obtained from our observations, with the goal of characterizing the differences and similarities between the spatial distributions of the various line emissions.  In W44, the PCA reveals that the observed emission lines fall into five distinct groups, each of which exhibits a distinct spatial distribution : (1) lines of S and H$_2\,(J > 2)$; (2) the H$_2$ S(0) line; (3) lines of ions with appearance potentials less than 13.6~eV; (4) lines of ions with appearance potentials greater than 13.6~eV, not including S$^{++}$; (5) lines of S$^{++}$.  In the other sources, some of these groups share a common distribution, but in no case do the spectral lines {\it within} a group defined above fail to be strongly correlated with each other.  This behavior suggests that the line emissions we have observed originate in at least five physically distinct components (some of which have similar distributions -- as projected onto the sky -- in sources other than W44).  

\noindent 4) The observed lines of S and H$_2\,(J > 2)$ likely originate in molecular material subject to a slow, nondissociative shock that is driven by the overpressure within the supernova remnant.  For the case of IC443C, where the gas pressure is constrained by the HD R(4)/R(3) ratio (Neufeld et al.\ 2006b), we estimate the atomic sulfur abundance as $\sim 10^{-6}$ relative to H$_2$ \re{(with a large uncertainty)}, a value corresponding to $\sim 4\%$ of the solar abundance of sulfur.  \re{Given the possible presence of S$^+$ in the [SI] emitting-region, this observed atomic sulfur abundance might be entirely consistent with a negligible depletion of sulfur onto grains and a negligible fraction of the sulfur nuclei being bound within gas-phase molecules.} 

\noindent 5) The H$_2$ S(0) line is typically more extended than any of the other spectral lines we have observed. In W28, it shows clear evidence for both a shock-excited component and a low-density diffuse emission component.

\noindent 6) The fine structure emissions from singly- and doubly-charged ions originate primarily in dissociative shock fronts with a range of shock velocities. Those ions with appearance potentials greater than 13.6~eV originate in faster shocks than -- and show a different distribution from --  ions with appearance potentials less than 13.6~eV.  In W44 the S$^{++}$ emissions show a very different distribution from those of other ions of similar appearance potential (e.g. Ne$^+$), a behavior which may reflect the lower critical density for the observed  S$^{++}$ emission lines.

\acknowledgments

\re{This work, which was supported in part by RSA agreement 1263841, is based on observations made with the {\it Spitzer Space Telescope}, which is operated by the Jet Propulsion Laboratory, California Institute of Technology, under a NASA contract.
D.A.N.\ and P.S.\ gratefully acknowledge the additional support of grant NAG5-13114 from NASA's Long Term Space Astrophysics (LTSA) Research Program.  We thank the anonymous referee for a very careful reading of the manuscript and a number of valuable suggestions.}

\clearpage

\begin{deluxetable}{cccclc}
\tablewidth{0pt}
\tablecaption{Details of the observations}
\tablehead{Source & Date & Module & Observing  & \re{Map center position in equatorial (J2000)} & Map size \\
                  &      &        &  time$^a$ (s) &  \re{and Galactic (decimal degrees) coordinates } & (arcsec) \\   }
\tabletypesize{\scriptsize}
\startdata
W44	&  2005 Apr 24    & SL &  5552 & $\rm (R.A.,Dec.) = (18h56m28.4s,+01^\circ 29^\prime59.0^{\prime \prime})$&  
$58 \times 57$         \\
W44	&  2005 Apr 24    & SH &  6625 & $\rm (l,b) = (34.8407,-0.4361)$ &  
$57 \times 58$         \\
W44	&  2005 Oct 11    & LH &  2650 &  &  $59 \times 56$         \\
\hline
W28	&  2005 Apr 19    & SL &  5552 & $\rm (R.A.,Dec.) = (18h01m52.3s,-23^\circ 19^\prime26.0^{\prime \prime})$ &  
$58 \times 57$         \\
W28	&  2005 Apr 19    & SH &  6625 & $\rm (l,b) = (6.6856,-0.2956)$ &  $57 \times 58$         \\
W28	&  2005 Apr 19    & LH &  2650 &  &  $59 \times 56$         \\
\hline
3C391	&  2005 Apr 17    & SL &  5552 & $\rm (R.A.,Dec.) = (18h49m21.9s,-00^\circ 57^\prime22.0^{\prime \prime})$ &  $58 \times 57$         \\
3C391	&  2005 Apr 17    & SH &  6625 & $\rm (l,b) = (31.8447,0.0252)$ &  $57 \times 58$         \\
3C391	&  2005 Apr 18    & LH &  2650 &  &  $59 \times 56$         \\
\hline
IC443C	&  2005 Mar 14    & SL &  5552 & $\rm (R.A.,Dec.) = (06h17m42.8s,22^\circ 21^\prime29.0^{\prime \prime})$  &  $58 \times 57$         \\
IC443C	&  2005 Mar 14    & SH &  6625 & $\rm (l,b) = (189.2953,3.0250)$  &  $57 \times 58$         \\
IC443C	&  2005 Mar 14    & LH &  2650 &  &  $59 \times 56$         \\
\hline

\enddata
\tablenotetext{a}{One cycle was used in each case, with a 60 sec ramp time for LH and SL, and a 30 sec ramp time for SH.}
\end{deluxetable}

\begin{deluxetable}{lrcccc}
\tablewidth{0pt}
\tablecaption{Beam-averaged line intensities, in a 25$^{\prime\prime}$ (HPBW) Gaussian beam} 
\tablehead{
Transition & Wavelength &\multicolumn{4}{c}{Intensity ($\rm 10^{-6}\,erg \, cm^{-2} \, s^{-1} \, sr^{-1}$)}\\
\cline{3-6}
& ($\mu$m)&W44 & W28 & 3C391 & IC443C} 
\tabletypesize{\scriptsize}
\startdata
H$_2$ S(0)  $\,\,\,\,J=2-0$ 			     & 28.2188	& 7.6	& 36.2	& 12.7	& 12.9 \\
H$_2$ S(1)  $\,\,\,\,J=3-1$                          & 17.0348	& 113	& 275	& 60.3	& 359  \\
H$_2$ S(2)  $\,\,\,\,J=4-2$                          & 12.2786 	& 135	& 392	& 81.7	& 499  \\
H$_2$ S(3)  $\,\,\,\,J=5-3$                          & 9.6649	& 324	& 622	& 151	& 1930 \\
H$_2$ S(4)  $\,\,\,\,J=6-4$                          & 8.0251	& 247	& 233	& 143	& 896  \\
H$_2$ S(5)  $\,\,\,\,J=7-5$                          & 6.9095	& 808	& 589	& 661	& 2331 \\
H$_2$ S(6)  $\,\,\,\,J=8-6$                          & 6.1086	&...$^a$&...$^a$&...$^a$& 491  \\
H$_2$ S(7)  $\,\,\,\,J=9-7$                          & 5.5112	& 436	& 231	& 272	& 1087 \\
HD R(3)	    $\,\,\,\,J=4-3$ 			     & 28.502   &	&       &       & 5.2 \\
HD R(4)	    $\,\,\,\,J=5-4$ 			     & 23.034	&       &       &       & 2.1 \\	
S $\,\,\,\rm ^3P_{1} -  ^3P_{2}$	             & 25.2490	& 1.7	& 16.2	& 26.8	& 30.3 \\
\hline
Si$^+$ $\,\,\,^2P^{\,0}_{3/2} - ^2P^{\,0}_{1/2}$     & 34.8141 	& 149		& 415	& 835	& 97.3 	\\
P$^+$$\,\,\,\rm ^3P_{2} -  ^3P_{1}$		     & 32.8709	& 1.7		& 4.4	& 11.9  &  $2.8 \pm 0.4^b$  \\
Cl$^+$ $\,\,\,\rm ^3P_{1} -  ^3P_{2}$		     & 14.3678  & $1.6 \pm 0.5$	& 3.3	& 8.6   & $< 1.2 \,(3\sigma)$   \\
Fe$^+$ $\,\,\,\rm ^4F_{9/2} - ^6D_{9/2}$             &  5.3403  & 107   	& 151	& 388	& $31.6 \pm 5.0$ 	\\
Fe$^+$ $\,\,\,\rm ^4F_{7/2} - ^4F_{9/2}$             & 17.9363  & 4.7   	& 8.7	& 42.0	& 2.5	\\
Fe$^+$ $\,\,\,\rm ^4F_{5/2} - ^4F_{7/2}$             & 24.5186  & $0.7 \pm 0.12$& 2.5   & 12.5  & $< 1.2 \,(3\sigma)$\\
Fe$^+$ $\,\,\,\rm ^6D_{7/2} - ^6D_{9/2}$             & 25.9882  & 22.4  	& 44.7 	& 160.4 & 11.3	\\
Fe$^+$ $\,\,\,\rm ^6D_{5/2} - ^6D_{7/2}$             & 35.3491  & 5.3   	& 11.6	& 47.2	& 10.3	\\ 
\hline
Ne$^+$ $\,\,\,^2P^{\,0}_{1/2} - ^2P^{\,0}_{3/2}$     & 12.8149  & 156		& 247	& 538	& 15.5	\\
Ne$^{++}$ $\,\,\,\rm ^3P_{1} -  ^3P_{2}$	     & 15.5545	& 17.1		& 38.4	& 202	&  4.2	\\		
Ne$^{++}$ $\,\,\,\rm ^3P_{0} -  ^3P_{1}$	     & 36.0135	& $<2.6\,(3\sigma)$&$3.4 \pm 1.1$& 18.0	& $< 0.9 \,(3\sigma)$\\
S$^{++}$ $\,\,\,\rm ^3P_{2} -  ^3P_{1}$		     & 18.713	& 34.4		& 40.4	& 76.9	& $< 1.4 \,(3\sigma)$\\
S$^{++}$ $\,\,\,\rm ^3P_{1} -  ^3P_{0}$		     & 33.480	& 43.7		& 74.2	& 140	& $< 4 \,(3\sigma)$\\
Fe$^{++}$  $\,\,\,\rm ^5D_{3} -  ^5D_{4}$	     & 22.925	& $1.2 \pm 0.2$ & 3.5	& 7.1	& $< 0.5 \,(3\sigma)$	\\

\enddata
\tablenotetext{a}{The H$_2$~S(6) line intensity cannot be determined reliably, because the line is blended with a strong 6.2~$\mu$m PAH feature.}
%\tablenotetext{b}{From Neufeld et al.\ (2006b)}
\tablenotetext{b}{Error bars, where given, are $1\sigma$ statistical errors.  When error bars are not given, the uncertainty is dominated by systematic errors, which we estimate to be $\le 25\%$ (see Neufeld et al.\ 2006a).}
\end{deluxetable}

\begin{deluxetable}{lcccc}
\tablewidth{0pt}
\tablecaption{H$_2$ parameters}
\tablehead{
\cline{2-5}
&W44 & W28 & 3C391 & IC443C} 
\tabletypesize{\scriptsize}
\startdata
Rotational state &  \multicolumn{4}{c}{log$_{10}\,$(Column density in cm$^{-2}$)}\\
J=2	& 19.73 & 20.36	& 19.97	& 19.98\\
J=3	& 19.52	& 19.83	& 19.28	& 20.06\\
J=4	& 18.71	& 19.09	& 18.54	& 19.32\\
J=5	& 18.63	& 18.70	& 18.41	& 19.52\\
J=6	& 17.79	& 17.69	& 17.60	& 18.40\\
J=7	& 17.84	& 17.66	& 17.78	& 18.32\\
J=8	&...$^{a}$&...$^{a}$&...$^{a}$& 17.33\\
J=9	& 16.96	& 16.63	& 16.79	& 17.39\\
\hline
\multicolumn{4}{l}{Two-parameter fits}\\
$T_w$ (K)                     & 347	& 322	& 266	&  617 	\\
log$_{10} \, (N_w$/cm$^{-2})$ & 20.39  	& 20.94	& 20.56	&  20.67\\
OPR$_w$                       & 1.58	& 0.93	& 0.65	&  2.42	\\
\\
$T_h$ (K)                       &1134	& 1039	& 1099	&  1225 	\\
log$_{10} \, (N_h$/cm$^{-2})$   &19.42	& 19.51	& 19.31	&  19.60	\\
OPR$_h$                         &...$^{a}$& ...$^{a}$& ...$^{a}$	&  3.85	\\
\hline
log$_{10} \, [(N_w + N_h$)/cm$^{-2}$]  & 20.44	& 20.95	& 20.58	& 20.70  	\\

\enddata
\tablenotetext{a}{The H$_2$~S(6) line intensity cannot be determined reliably, because the line is blended with a strong 6.2~$\mu$m PAH feature.  In these cases, the OPR cannot be determined for the hot component.}
\end{deluxetable}

\clearpage

\begin{figure}
\includegraphics[scale=0.65,angle=0]{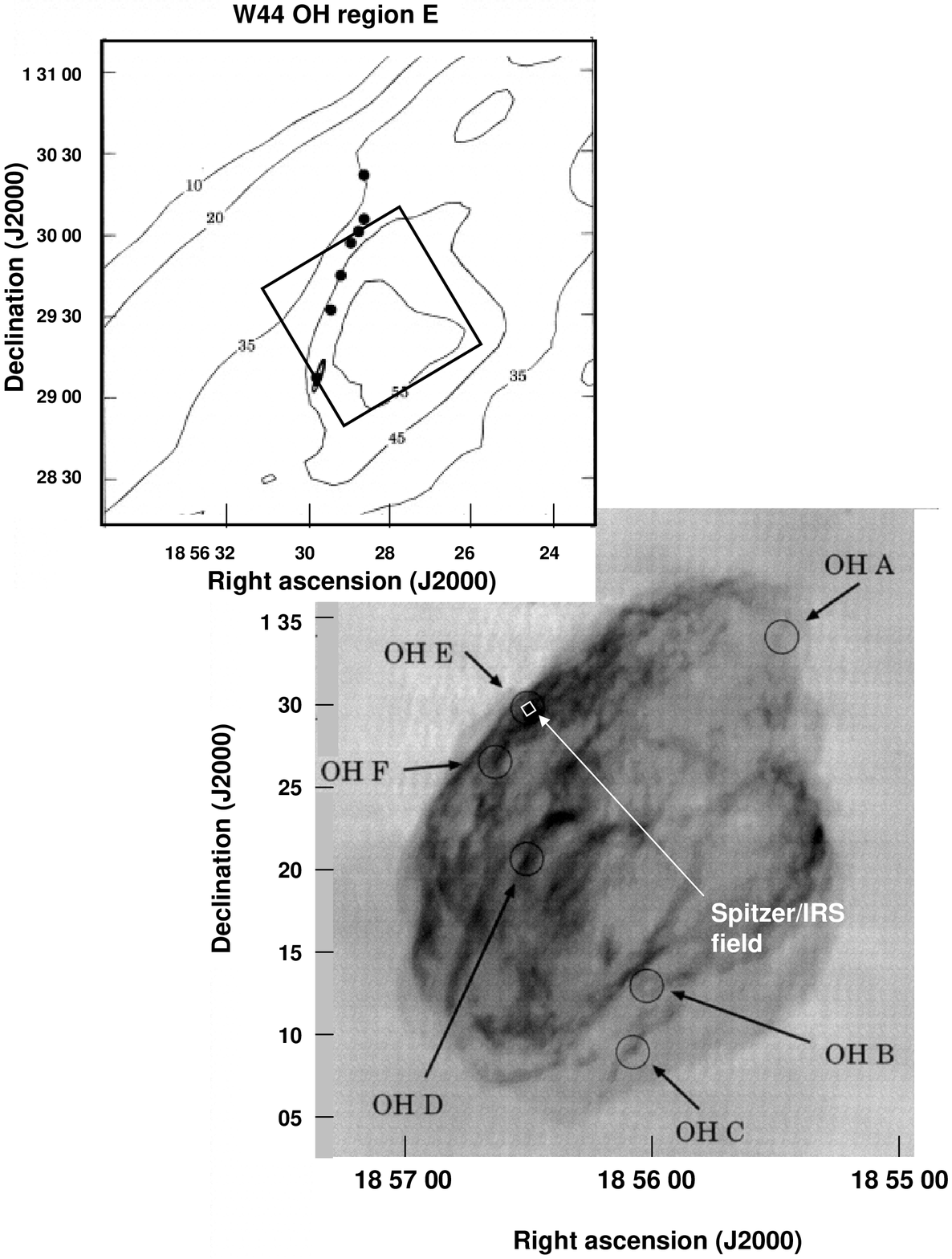}

\noindent{Fig.\ 1  -- Finder chart for W44.  The region mapped by the SH module \re{(black square in upper panel: white square with arrow in lower panel)} is shown relative to the OH maser emission and radio continuum emission in W44.  Upper left panel: the region mapped by the SH module is marked on Figure 5e of Claussen et al.\ (1997), which shows the location of the OH maser spots within W44 OH Region E as well as the 1720 MHz radio continuum intensity (contours, labeled in mJy/beam).  Lower right panel (from Claussen et al.\ 1997; their Figure 2): the location of the OH maser regions is shown on the 1442 MHz radio continuum image of Giacani et al.\ (1997).}  For an assumed distance to the source of 2.5 kpc, the map in the lower right panel has a linear size $\sim 25 \times 25$~pc, and the region mapped by {\it Spitzer} has a linear size $\sim 0.7 \times 0.7$~pc.
\end{figure}
\clearpage

\begin{figure}
\includegraphics[scale=0.65,angle=0]{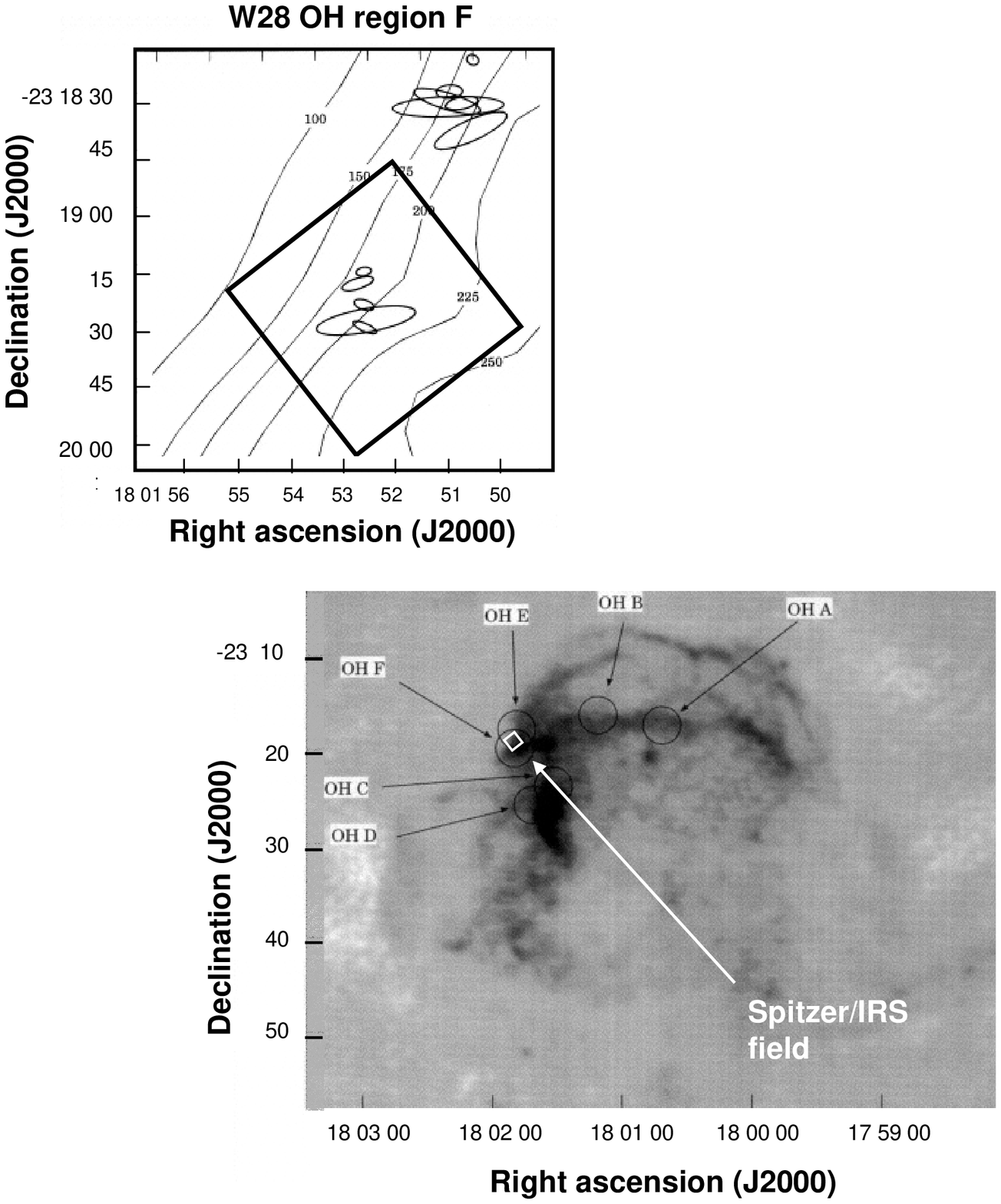}

\noindent{Fig.\ 2 -- Finder chart for W28.  The region mapped by the SH module \re{(black square in upper panel: white square with arrow in lower panel)} is shown relative to the OH maser emission and radio continuum emission in W28.  Upper left panel: the region mapped by the SH module is marked on Figure 4f of Claussen et al.\ (1997), which shows the location of the OH maser spots within W28 OH Region F as well as the 1720 MHz radio continuum intensity (contours, labeled in mJy/beam).  Lower right panel (from Claussen et al.\ 1997; their Figure 1): the location of the OH maser regions is shown on the 327 MHz radio continuum image of Frail et al.\ (1993).  For an assumed distance to the source of 1.9 kpc, the map in the lower right panel has a linear size $\sim 40 \times 30$~pc, and the region mapped by {\it Spitzer} has a linear size $\sim 0.55 \times 0.55$~pc.}
\end{figure}
\clearpage

\begin{figure}
\includegraphics[scale=0.75,angle=0]{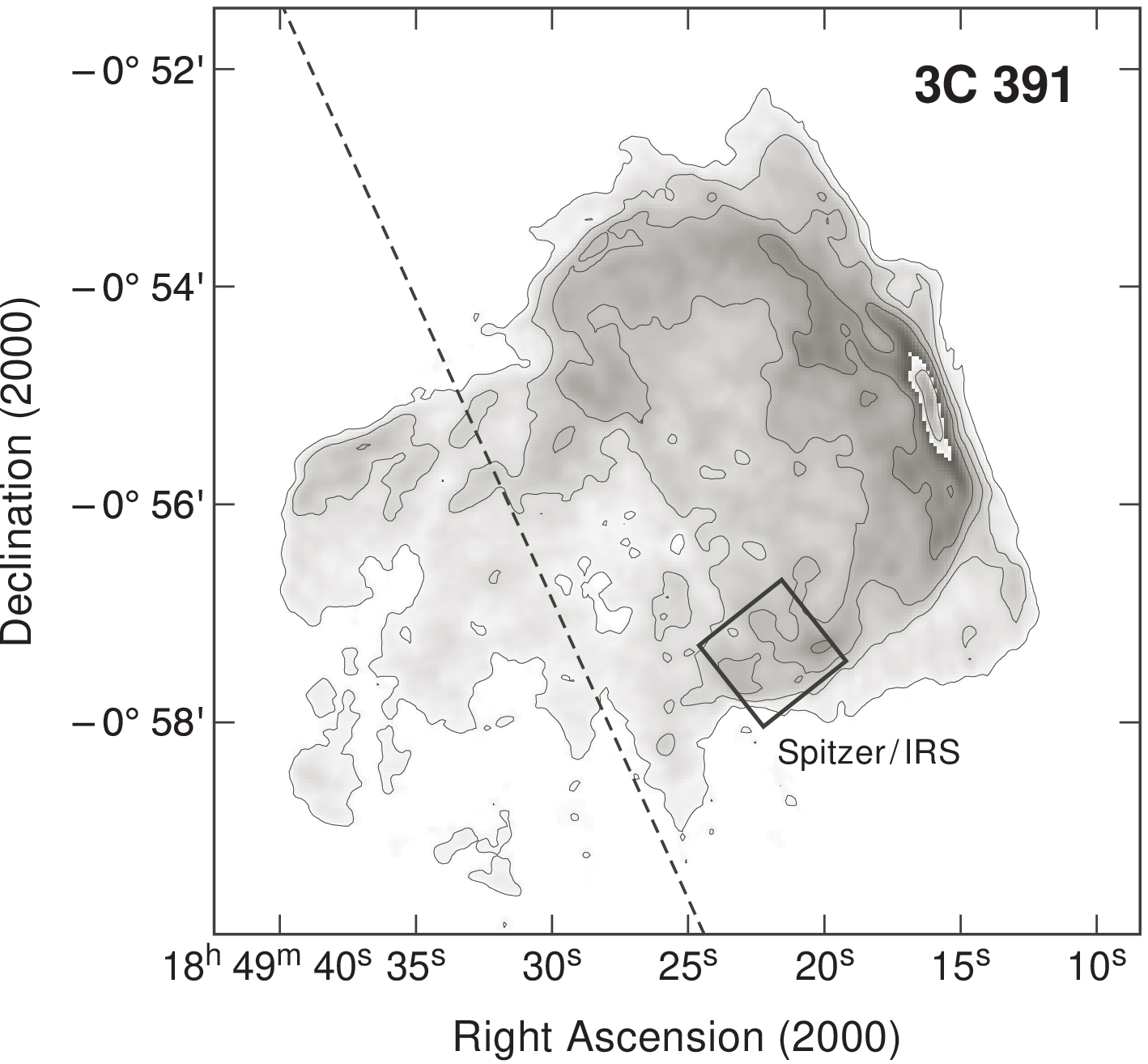}

\noindent{Fig.\ 3 -- Finder chart for 3C391.  Radio continuum (1446~MHz) image of 3C$\:$391 obtained
with a spatial resolution of 6\asec\ (from Reynolds \& Moffett 1993). The
Galactic plane is indicated by the dashed diagonal line. The box
centered on the ``broad molecular line region" at $\alpha=18^{\rm h}$ 49$^{\rm m}$ 21.$^{\rm \!\! s}$9, 
$\delta=-0^{\rm o}$ 57\amin\ 22\asec\ (J2000) delineates the area
mapped with the SH module.  For an assumed distance to the source of 9 kpc, the entire map has a linear size 
$\sim 20 \times 20$~pc, and the region mapped by {\it Spitzer} has a linear size $\sim 2.5 \times 2.5$~pc. }
\end{figure}
\clearpage

\begin{figure}
\includegraphics[scale=0.60,angle=0]{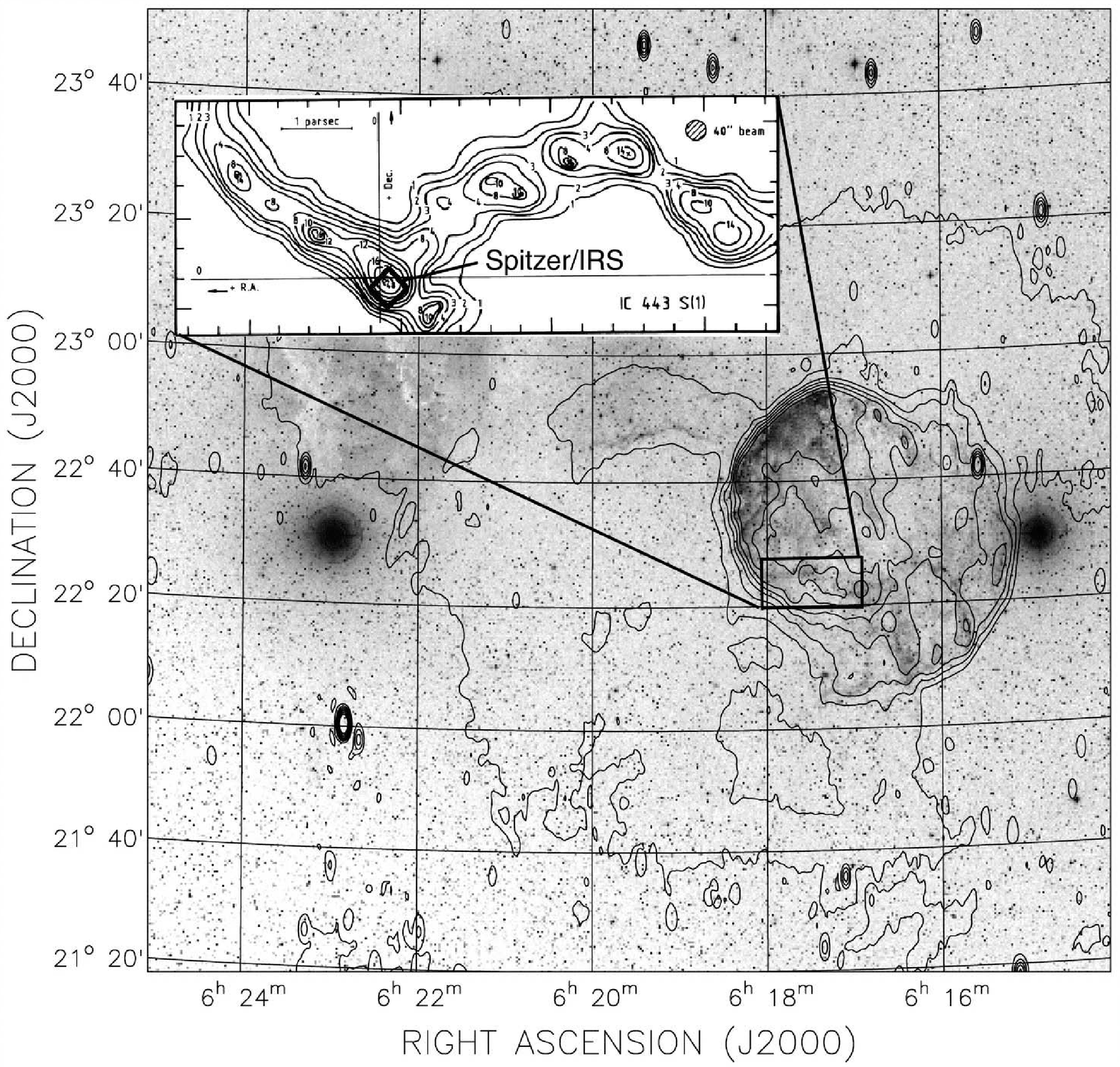}

\noindent{Fig.\ 4 -- Finder chart for IC443.  An optical image of IC443 with
contours of the 1420 MHz continuum emission overlaid (Leahy 2004).
The insert shows an expanded region of the remnant with the distribution 
of H$_2$ $v = 1-0$ emission obtained by Burton et al. (1988).  The region 
mapped by the SH module (black square) is shown in the insert overlaid
on the contours of the H$_2$ emission.   For an assumed distance to the source of 1.5 kpc, the insert has a linear size $\sim 9 \times 3.5$~pc, and the region mapped by {\it Spitzer} has a linear size $\sim 0.45 \times 0.45$~pc.}  
\end{figure}
\clearpage

\begin{figure}
\includegraphics[scale=0.75,angle=0]{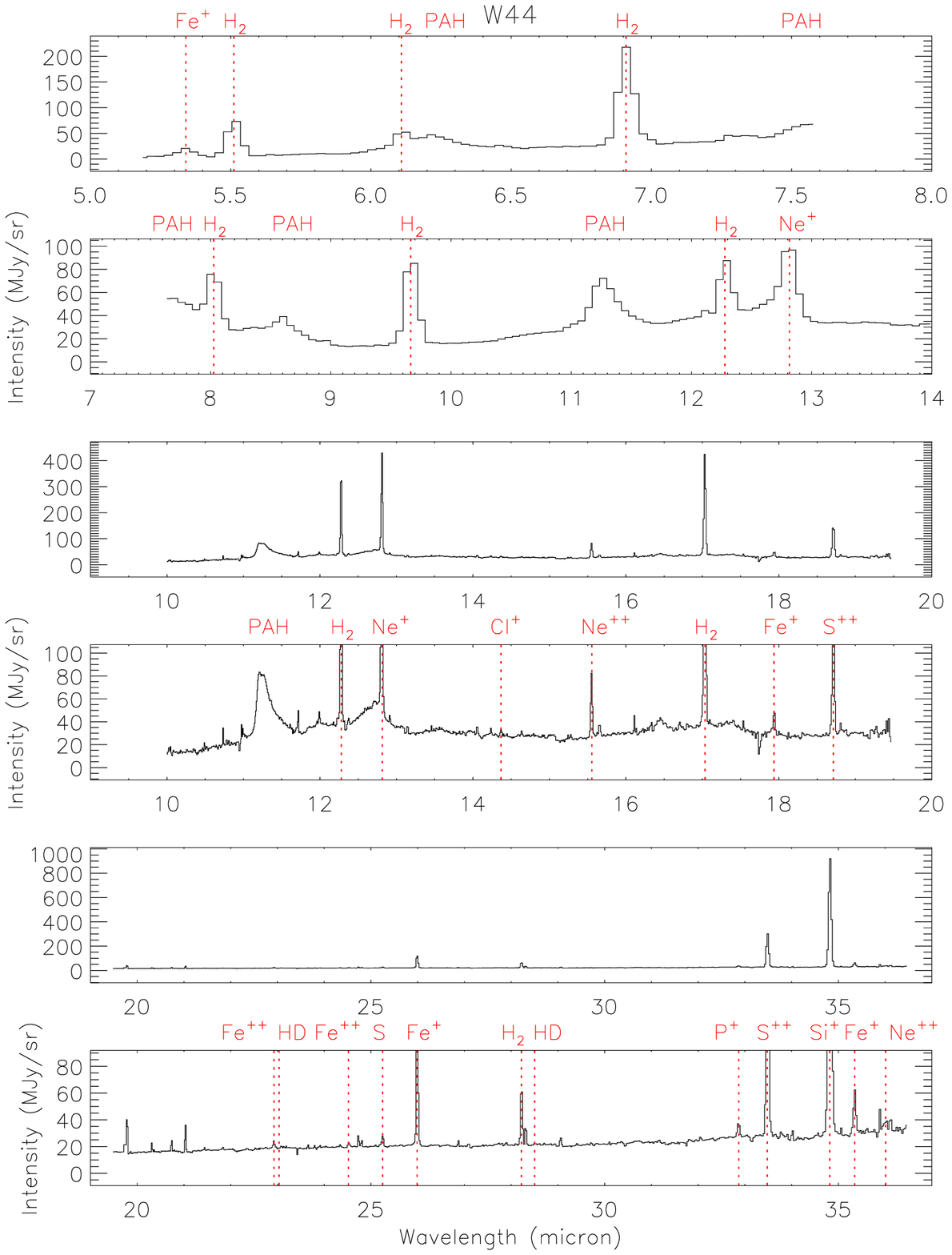}

\noindent{Fig.\ 5 -- Average spectra observed for a $25^{\prime\prime}$ (HPBW) diameter circular aperture centered at 
$\alpha=18^{\rm h}$ 56$^{\rm m}$ 28.$^{\rm \!\! s}$4, 
$\delta=+0^{\rm o}$ 29\amin\ 59\asec\ (J2000)
in W44 (corresponding to offset $(0^{\prime\prime},0^{\prime\prime})$ in Figure 10 below).  The spectra in the upper two panels were obtained with the SL module, those in the middle two panels were obtained with the SH module ($\lambda \le \rm 19.5\,\mu m$) and those in the lower two panels with the LH module ($\lambda \ge \rm 19.5\,\mu m$).  (For each of the SH and LH spectra, two panels are shown with different vertical scales.)
Vertical dotted lines demark the wavelengths of spectral lines listed in Table 2 (regardless of whether the line is actually detected in this particular source.)  \re{The spectra shown here are not background-subtracted, since no off-source measurements were made, and thus the continuum flux level must be regarded as somewhat uncertain.}
}
\end{figure}

\begin{figure}
\includegraphics[scale=0.75,angle=0]{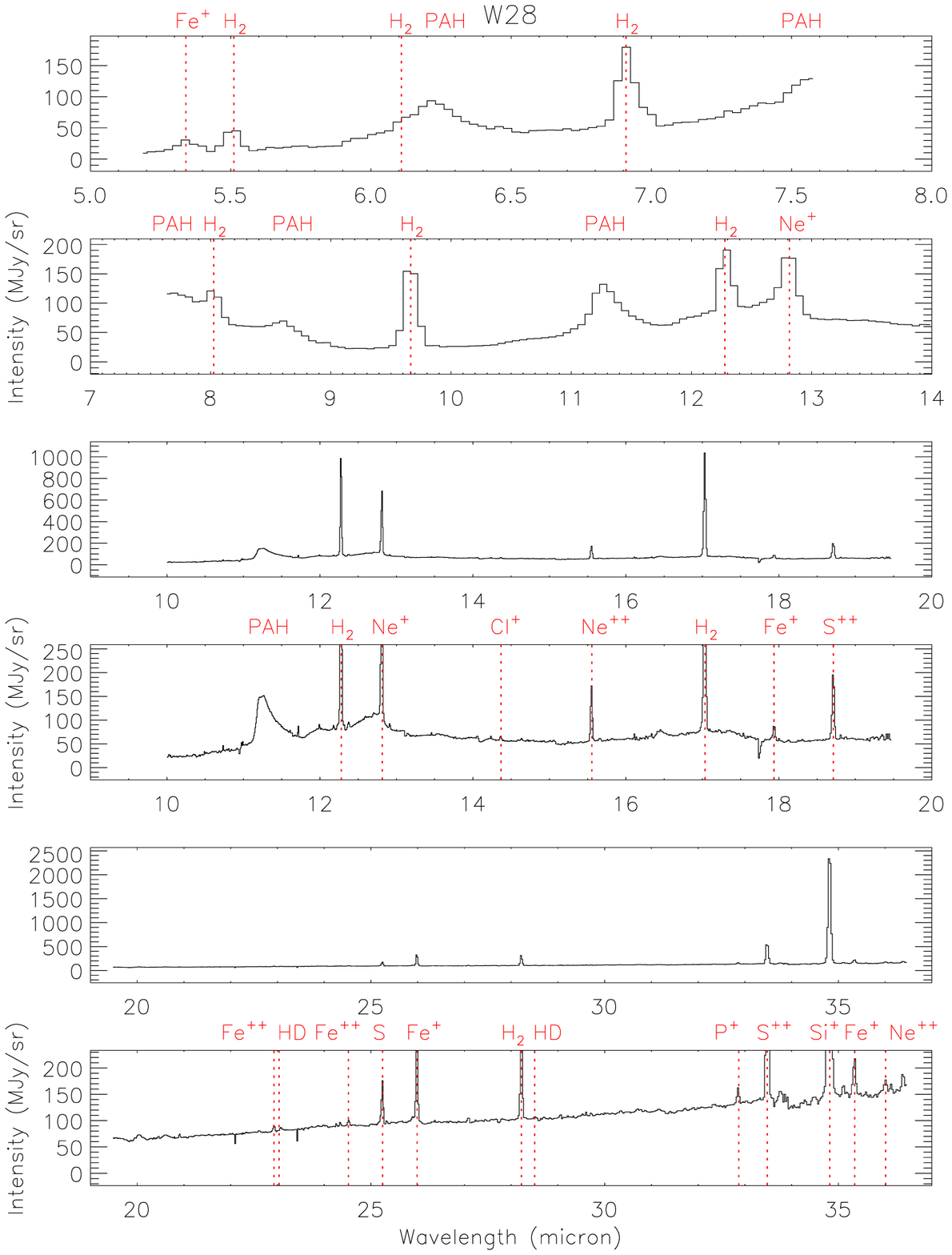}

%\noindent{Fig.\ 2b -- Average spectra observed for a $25^{\prime\prime}$ (HPBW) diameter circular aperture centered at $\alpha=$18h\,01m\,52.3s, $\delta=-23$d\,19$^\prime$25$^\prime$$^\prime$ in W28 (corresponding to offset $(0^{\prime\prime},0^{\prime\prime})$ in Figure 4b below).  The spectra in the upper two panels were obtained with the SL module, those in the middle two panels were obtained with the SH module ($\lambda \le \rm 19.5\,\mu m$) and those in the lower two panels with the LH module ($\lambda \ge \rm 19.5\,\mu m$).   Vertical dotted lines demark the wavelengths of spectral lines listed in Table 2 (regardless of whether the line is actually detected in this particular source.)  Red lines denote rotational transitions of H$_2$ or HD, while blue (or green for [FeII]) lines denote fine structure transitions.}
\noindent{Fig.\ 6 -- same as for Fig.\ 5, except for a $25^{\prime\prime}$ (HPBW) diameter circular aperture 
centered at 
$\alpha=18^{\rm h}$ 01$^{\rm m}$ 52.$^{\rm \!\! s}$3, 
$\delta=-23^{\rm o}$ 19\amin\ 25\asec\ (J2000)
in W28 (corresponding to offset $(0^{\prime\prime},0^{\prime\prime})$ in Figure 11 below).}
\end{figure}

\begin{figure}
\includegraphics[scale=0.75,angle=0]{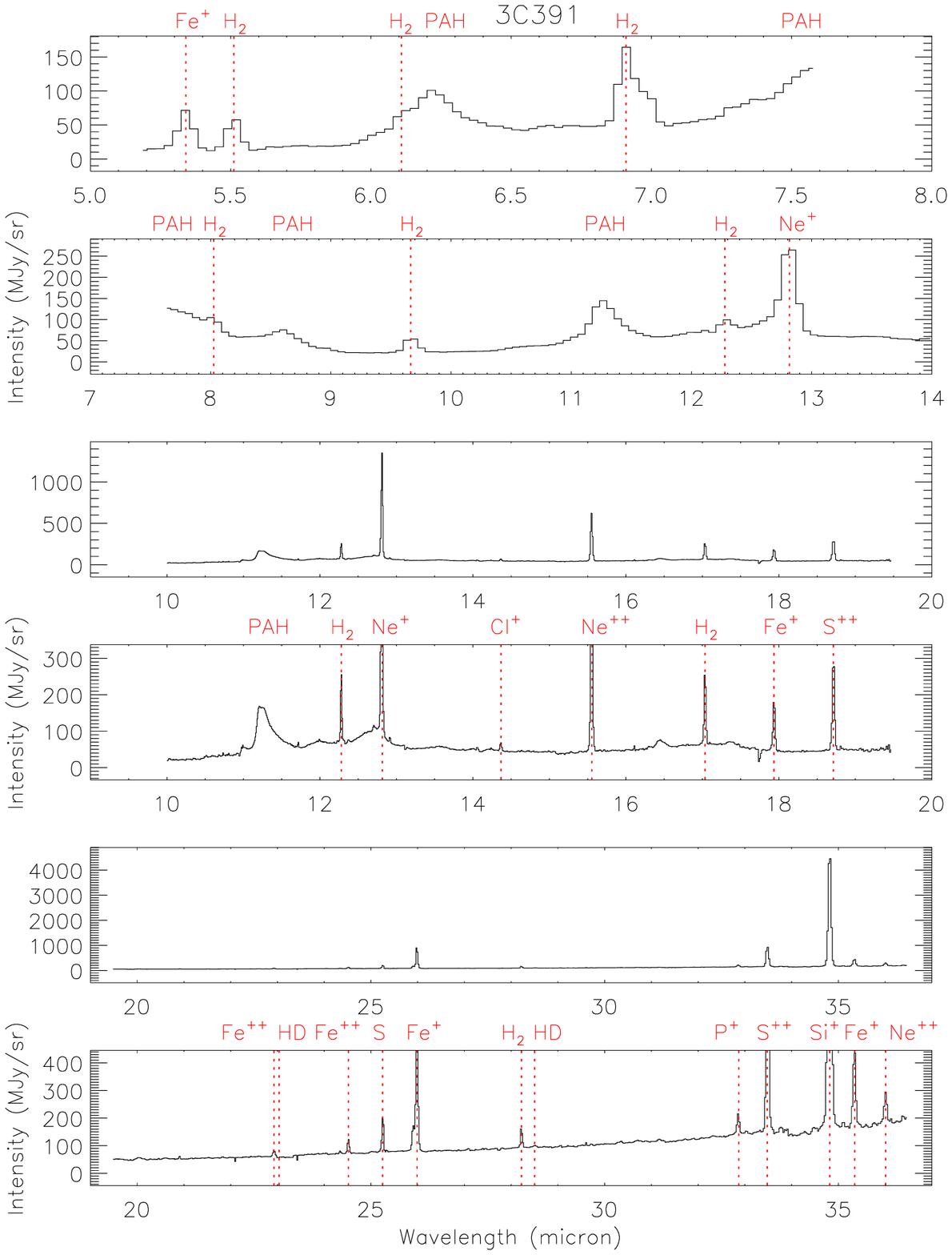}

%\noindent{Fig.\ 2c -- Average spectra observed for a $25^{\prime\prime}$ (HPBW) diameter circular aperture centered at $\alpha=$18h\,49m\,21.9s, $\delta=-00$d\,57$^\prime$22 OH $^\prime$$^\prime$ in 3C391 (corresponding to offset $(0^{\prime\prime},0^{\prime\prime})$ in Figure 4c below).  The spectra in the upper two panels were obtained with the SL module, those in the middle two panels were obtained with the SH module ($\lambda \le \rm 19.5\,\mu m$) and those in the lower two panels with the LH module ($\lambda \ge \rm 19.5\,\mu m$). Vertical dotted lines demark the wavelengths of spectral lines listed in Table 2 (regardless of whether the line is actually detected in this particular source.)  Red lines denote rotational transitions of H$_2$ or HD, while blue (or green for [FeII]) lines denote fine structure transitions.}
\noindent{Fig.\ 7 -- same as for Fig.\ 5, except for a $25^{\prime\prime}$ (HPBW) diameter circular aperture 
centered at 
$\alpha=18^{\rm h}$ 49$^{\rm m}$ 21.$^{\rm \!\! s}$9, 
$\delta=-0^{\rm o}$ 57\amin\ 22\asec\ (J2000) 
in 3C391 (corresponding to offset $(0^{\prime\prime},0^{\prime\prime})$ in Figure 12 below).}

\end{figure}

\begin{figure}
\includegraphics[scale=0.75,angle=0]{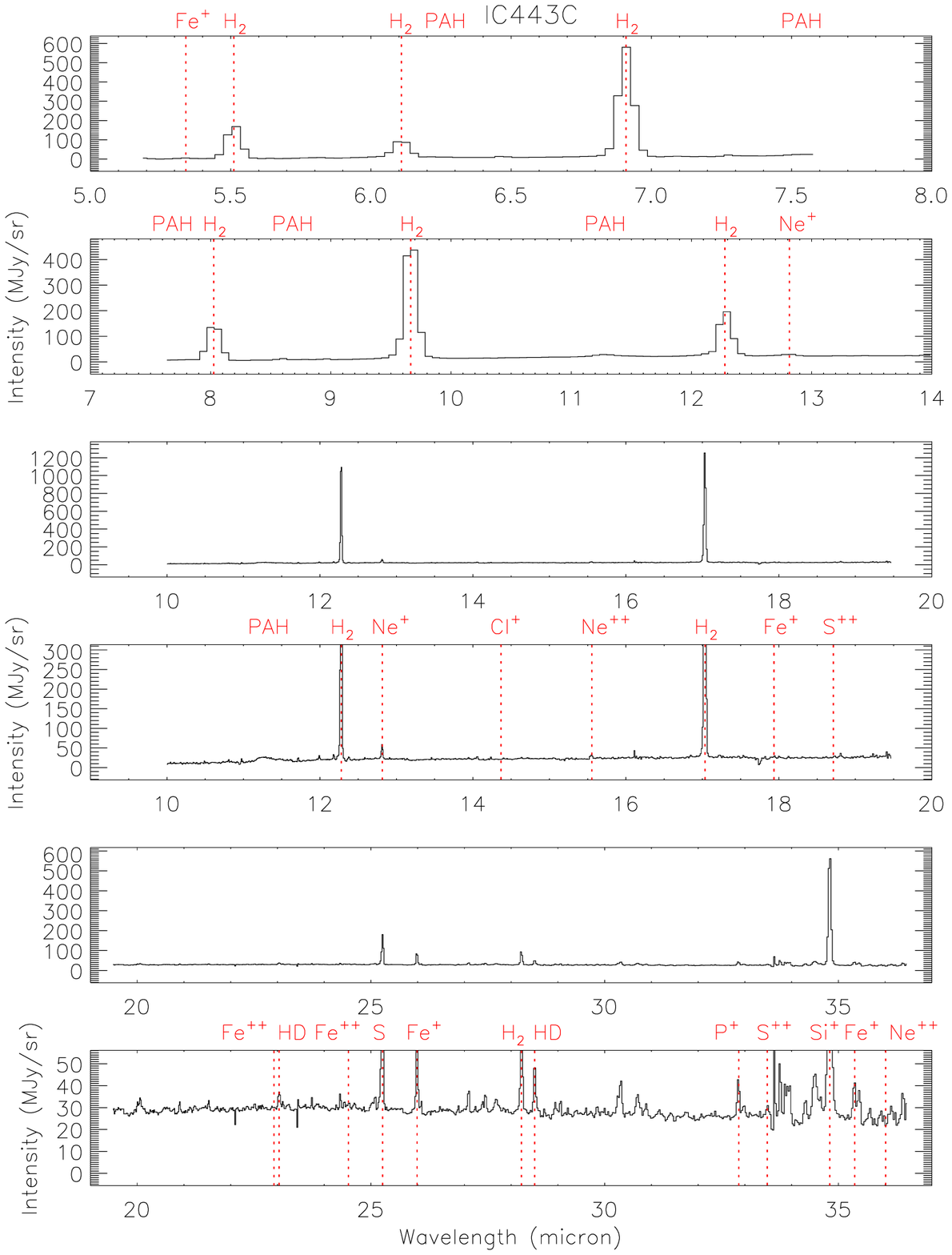}

%\noindent{Fig.\ 2d -- Average spectra observed for a $25^{\prime\prime}$ (HPBW) diameter circular aperture centered at $\alpha=$00h\,00m\,00.00s, $\delta=+00$d\,00$^\prime$00.0$^\prime$$^\prime$ in IC443C (corresponding to offset $(-24^{\prime\prime},-20^{\prime\prime})$ in Figure 4d below). The spectra in the upper two panels were obtained with the SL module, those in the middle two panels were obtained with the SH module ($\lambda \le \rm 19.5\,\mu m$) and those in the lower two panels with the LH module ($\lambda \ge \rm 19.5\,\mu m$). Vertical dotted lines demark the wavelengths of spectral lines listed in Table 2 (regardless of whether the line is actually detected in this particular source.)  Red lines denote rotational transitions of H$_2$ or HD, while blue (or green for [FeII]) lines denote fine structure transitions.}
\noindent{Fig.\ 8 -- same as for Fig.\ 5, except for a $25^{\prime\prime}$ (HPBW) diameter circular aperture 
centered at 
$\alpha=06^{\rm h}$ 17$^{\rm m}$ 42.$^{\rm \!\! s}$5, 
$\delta=+22^{\rm o}$ 21\amin\ 29\asec\ (J2000)
in IC443C (corresponding to offset $(-24^{\prime\prime},-20^{\prime\prime})$ in Figure 13 below).}

\end{figure}

\begin{figure}
\includegraphics[scale=0.75,angle=0]{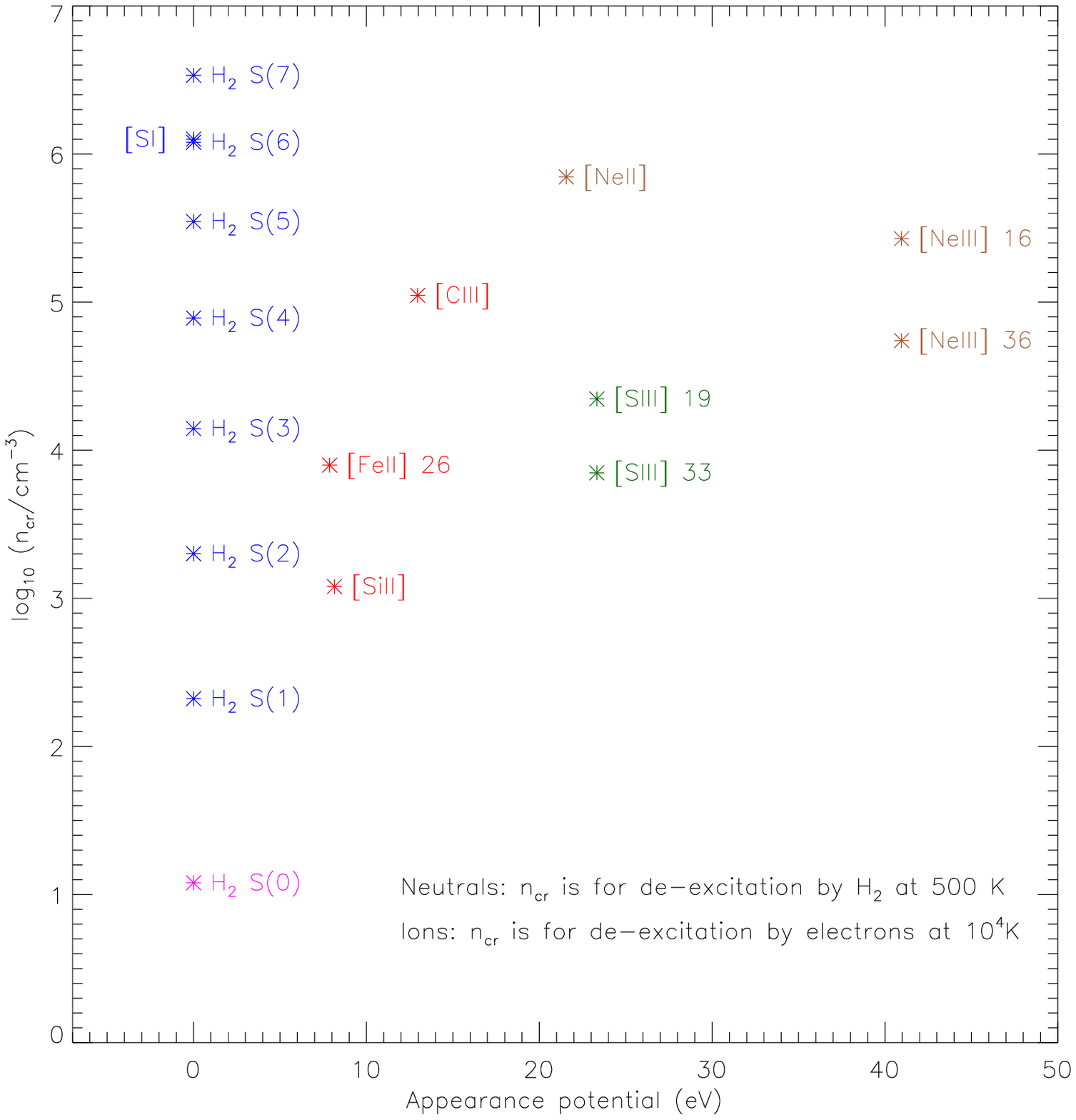}

\noindent{Fig.\ 9 -- Excitation conditions for the observed transitions.  The horizontal axis shows the appearance potential -- equal to zero for neutral species and equal to the ionization potential of X$^{(n-1)+}$ for the case of ion X$^{n+}$ -- and the vertical axis shows the critical density, $n_{\rm cr}$, at which the rate of collisional de-excitation equals the spontaneous radiative rate. 
For neutral species, the critical densities apply to de-excitation by H$_2$ molecules at a temperature of 500~K; for ionized species, the critical densities apply to de-excitation by electrons at a temperature of 10$^4$~K.  The
values adopted for $n_{\rm cr}$ are from the papers of Giveon et al.\ (2002), Le Bourlot, Pineau des For{\^e}ts \& Flower (1999) and HM89.}
\end{figure}

\begin{figure}
\includegraphics[scale=0.95,angle=0]{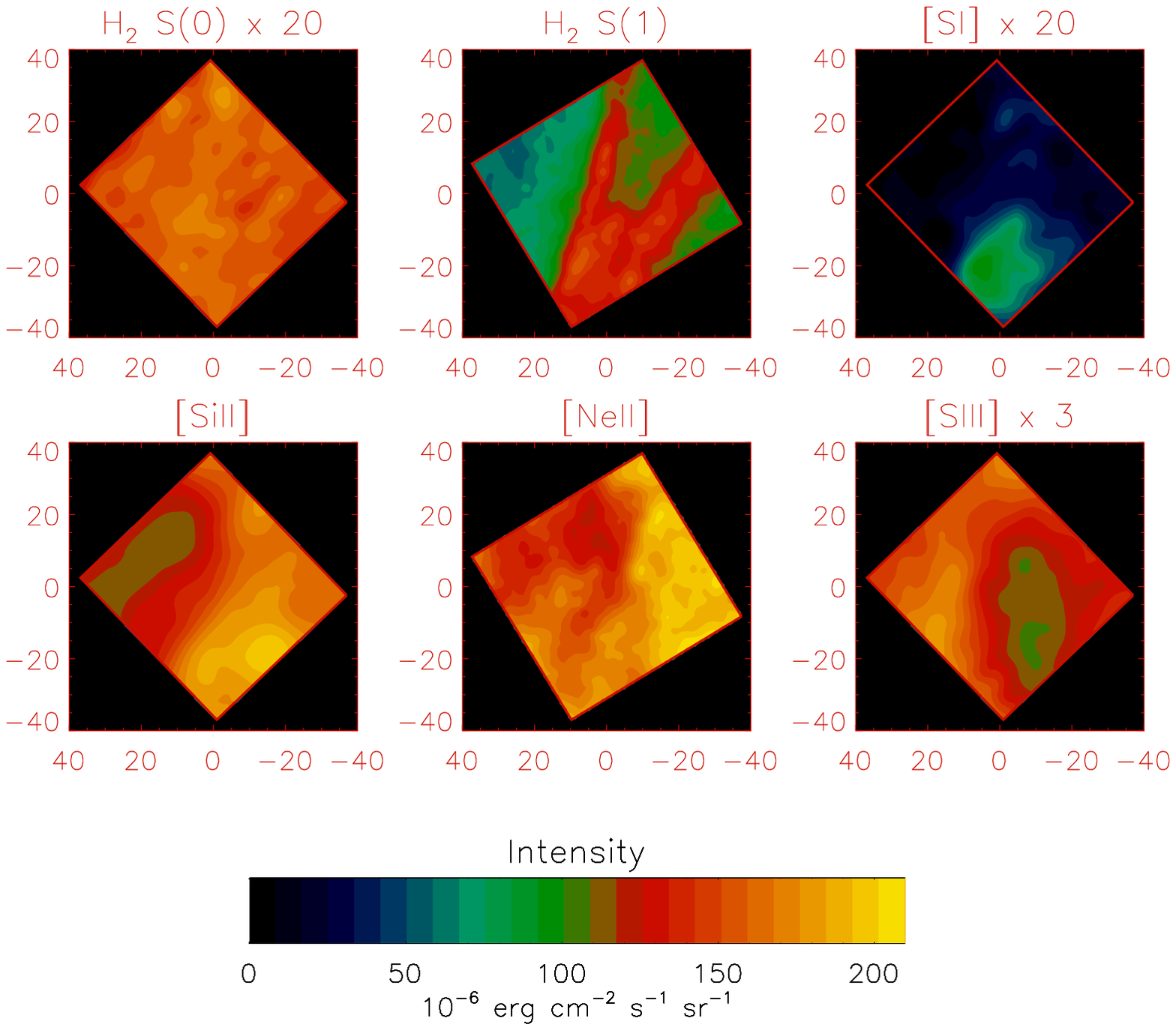}

\noindent{Fig.\ 10 -- H$_2$ S(0), H$_2$ S(1), [SI], [SiII], [NeII], and [SIII] emission line 
intensities observed 
toward W44.   The inset boxes demark the regions within which each transition was mapped.  The
horizontal and vertical axes show the R.A. ($\Delta \alpha \rm \cos \delta$) and declination ($\Delta
\delta$) offsets in arcsec relative to $\alpha=18^{\rm h}$ 56$^{\rm m}$ 28.$^{\rm \!\! s}$4, 
$\delta=+0^{\rm o}$ 29\amin\ 59\asec\ (J2000).}
\end{figure}
\clearpage

\begin{figure}
\includegraphics[scale=0.95,angle=0]{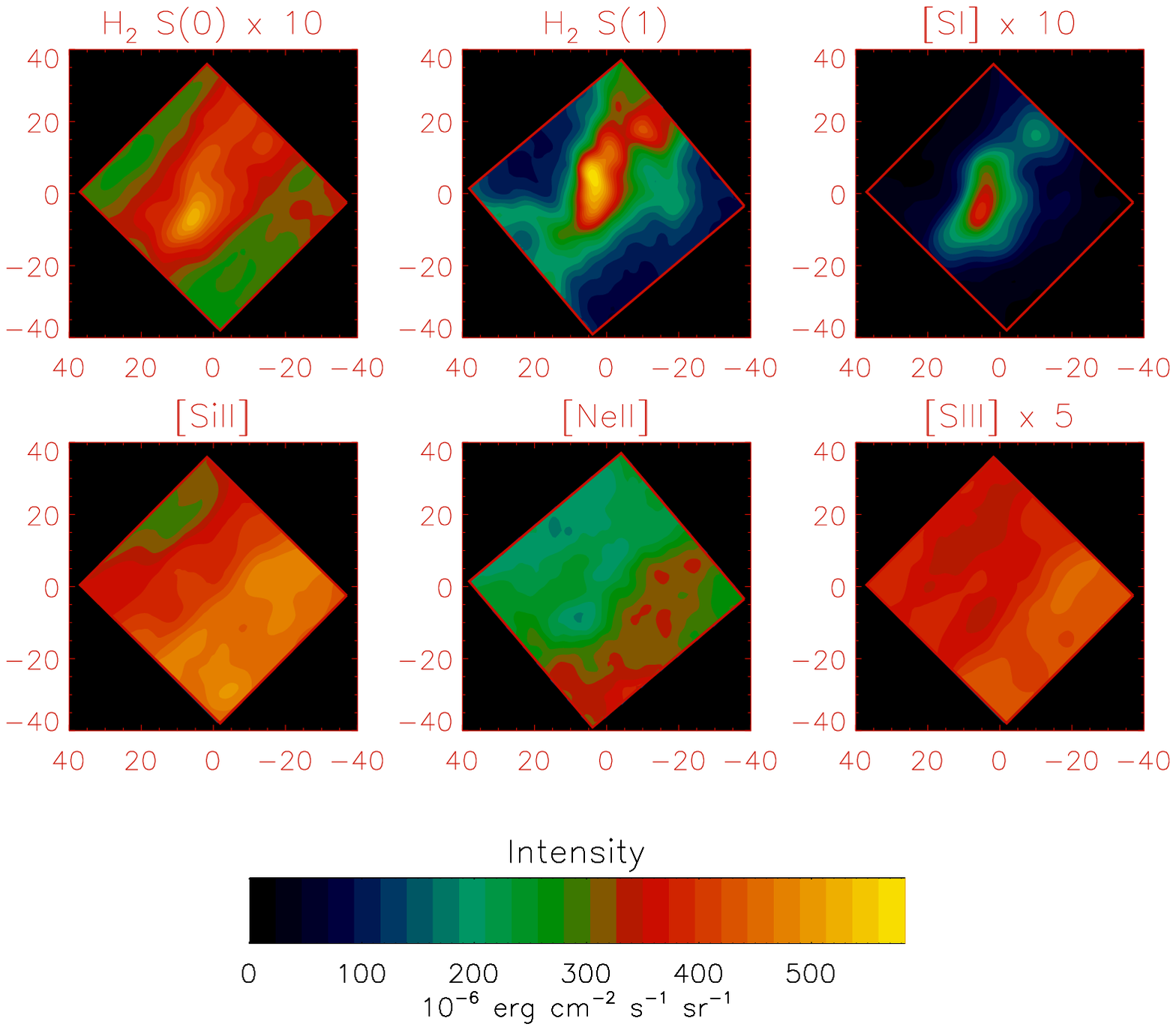}

\noindent{Fig.\ 11 -- H$_2$ S(0), H$_2$ S(1), [SI], [SiII], [NeII], and [SIII] emission line 
intensities observed 
toward W28.   The inset boxes demark the regions within which each transition was mapped.  The
horizontal and vertical axes show the R.A. ($\Delta \alpha \rm \cos \delta$) and declination ($\Delta
\delta$) offsets in arcsec relative to $\alpha=18^{\rm h}$ 01$^{\rm m}$ 52.$^{\rm \!\! s}$3, 
$\delta=-23^{\rm o}$ 19\amin\ 25\asec\ (J2000).}
\end{figure}
\clearpage

\begin{figure}
\includegraphics[scale=0.95,angle=0]{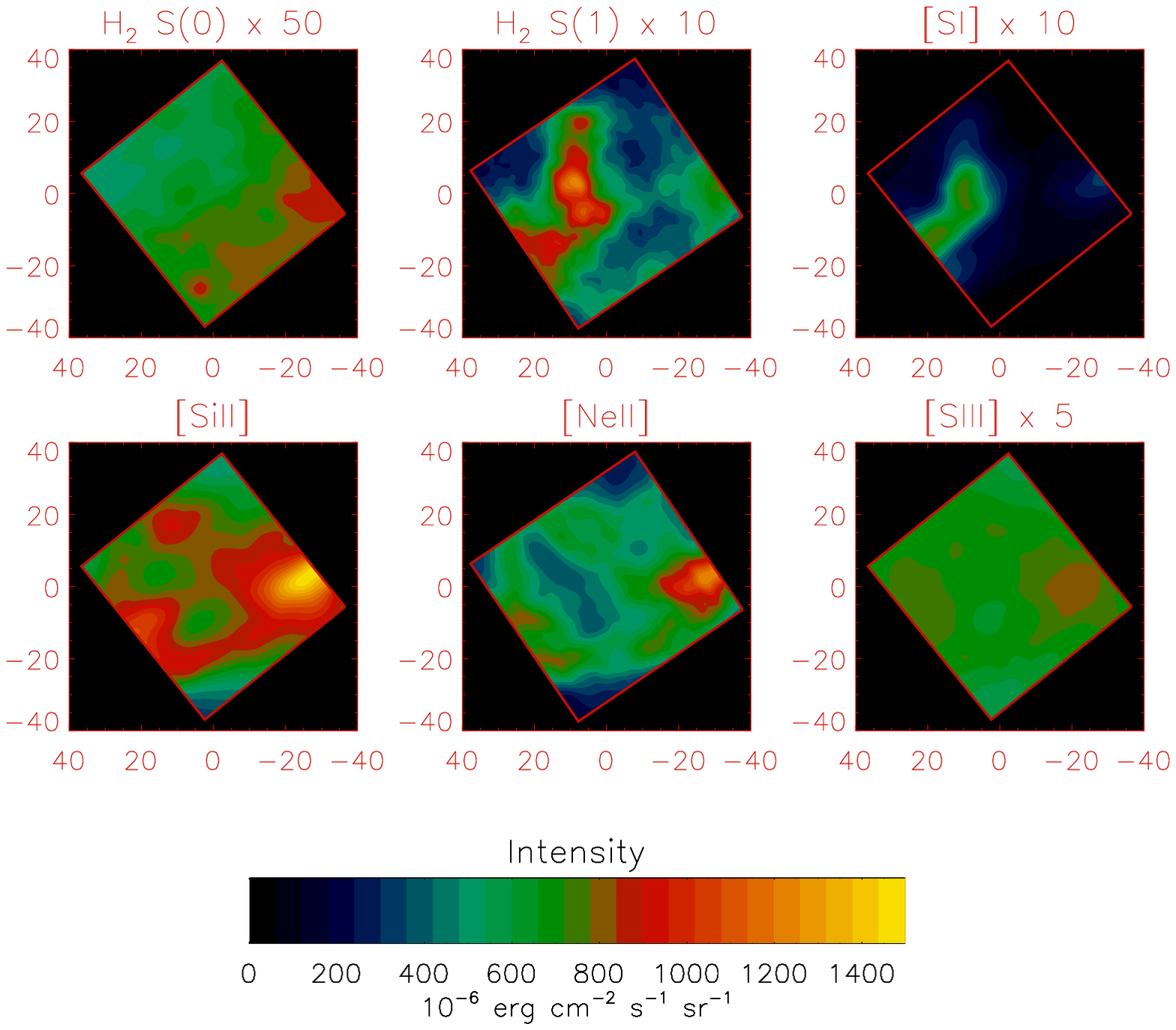}

\noindent{Fig.\ 12 -- H$_2$ S(0), H$_2$ S(1), [SI], [SiII], [NeII], and [SIII] emission line 
intensities observed toward 3C391.   The inset boxes demark the regions within which each transition was mapped.  The
horizontal
and vertical axes show the R.A. ($\Delta \alpha \rm \cos \delta$) and declination ($\Delta
\delta$) offsets in arcsec relative to $\alpha=18^{\rm h}$ 49$^{\rm m}$ 21.$^{\rm \!\! s}$9, 
$\delta=-0^{\rm o}$ 57\amin\ 22\asec\ (J2000).}
\end{figure}
\clearpage

\begin{figure}
\includegraphics[scale=0.95,angle=0]{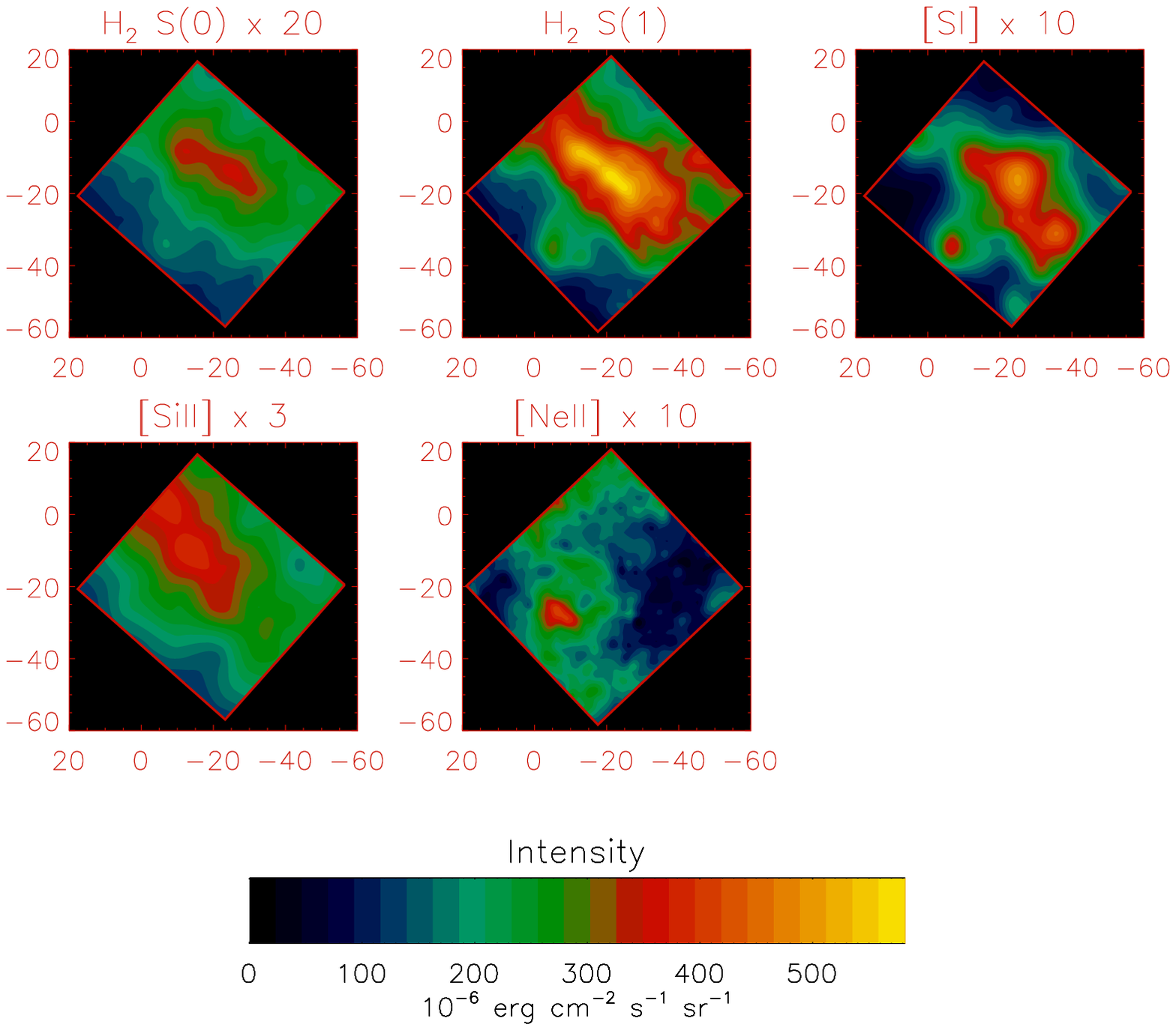}

\noindent{Fig.\ 13 -- H$_2$ S(0), H$_2$ S(1), [SI], [SiII], and [NeII] emission line 
intensities observed 
toward IC443C.   The inset boxes demark the regions within which each transition was mapped.  The
horizontal and vertical axes show the R.A. ($\Delta \alpha \rm \cos \delta$) and declination ($\Delta
\delta$) offsets in arcsec relative to 
$\alpha=06^{\rm h}$ 17$^{\rm m}$ 44.$^{\rm \!\! s}$2, 
$\delta=+22^{\rm o}$ 21\amin\ 49\asec\ (J2000).}
\end{figure}
\clearpage

\begin{figure}
\includegraphics[scale=0.75,angle=90]{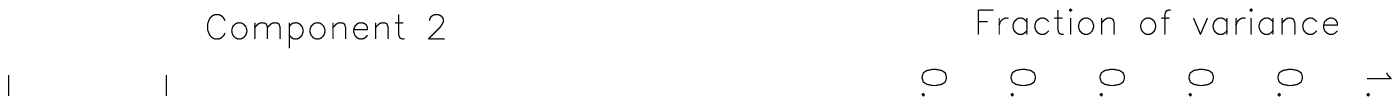}
\noindent{Fig. 14 -- Results of PCA analysis for W44.  Upper panel: fraction of the variance accounted for by each component (asterisks), along with the cumulative fraction for the first $n$ components as a function of $n$ (open squares).   Left lower panel: coefficients for the first and second principal components needed to approximate the maps of each transition.  Right lower panel: coefficients for the second and third
principal components.  Where multiple transitions of a single ion have been observed, the labels distinguish between the different transitions by giving the wavelength rounded to the nearest micron.}
\end{figure}
\clearpage

\begin{figure}
\includegraphics[scale=0.75,angle=90]{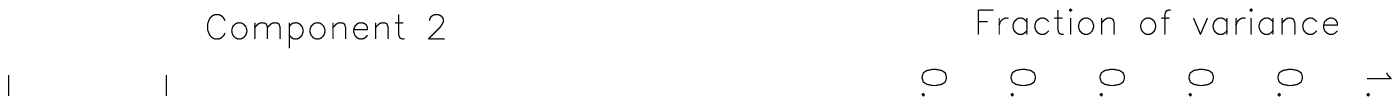}
%\noindent{Fig.\ 5b -- Results of PCA analysis for W28.  Upper panel: fraction the variance accounted for by each component (asterisks), along with the cumulative fraction for the first $n$ components as a function of $n$ (open squares).   Left lower panel: coefficients for the first and second principal components needed to approximate the maps of each transition.  Right lower panel: coefficients for the second and third principal components.  Where multiple transitions of a single ion have been observed, the labels distinguish between the different transitions by giving the wavelength rounded to the nearest micron.}
\noindent{Fig.\ 15 -- same as for Fig.\ 14, except for W28.} 
\end{figure}
\clearpage

\begin{figure}
\includegraphics[scale=0.75,angle=90]{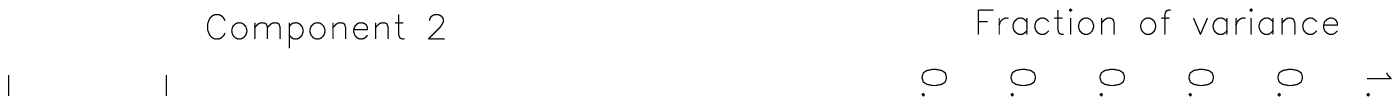}
%\noindent{Fig.\ 5c -- Results of PCA analysis for 3C391.  Upper panel: fraction of the variance accounted for by each component (asterisks), along with the cumulative fraction for the first $n$ components as a function of $n$ (open squares).   Left lower panel: coefficients for the first and second principal components needed to approximate the maps of each transition.  Right lower panel: coefficients for the second and third principal components.  Where multiple transitions of a single ion have been observed, the labels distinguish between the different transitions by giving the wavelength rounded to the nearest micron.}
\noindent{Fig.\ 16 -- same as for Fig.\ 14, except for 3C391.} 
\end{figure}
\clearpage

\begin{figure}
\includegraphics[scale=0.75,angle=90]{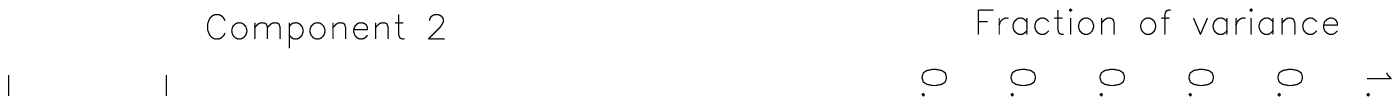}
%\noindent{Fig.\ 5d -- Results of PCA analysis for IC443C.  Upper panel: fraction of the variance accounted for by each component (asterisks), along with the cumulative fraction for the first $n$ components as a function of $n$ (open squares).   Left lower panel: coefficients for the first and second principal components needed to approximate the maps of each transition.  Right lower panel: coefficients for the second and third principal components.  Where multiple transitions of a single ion have been observed, the labels distinguish between the different transitions by giving the wavelength rounded to the nearest micron.}
\noindent{Fig.\ 17 -- same as for Fig.\ 14, except for IC443C.} 
\end{figure}
\clearpage

\begin{figure}
\includegraphics[scale=0.6,angle=0]{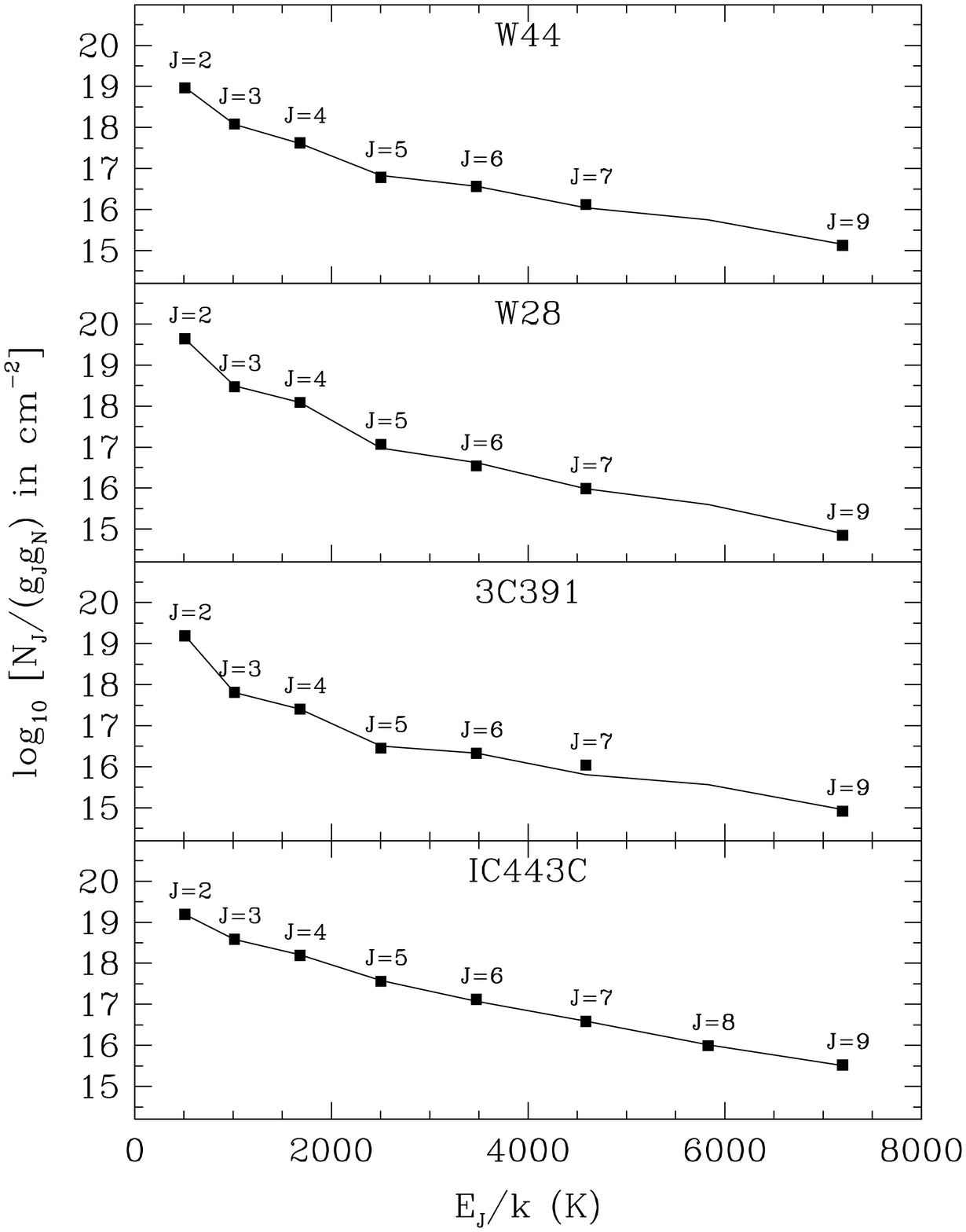}

\noindent{Fig.\ 18 -- H$_2$ rotational diagrams, computed for $25^{\prime\prime}$ (HPBW) diameter circular apertures centered at the positions given in the captions to Figures 5 -- 8.  \re{As usual, the quantity plotted on the vertical axis is the logarithm of the column density per magnetic substate, ${\rm log}_{10} (N / g_J g_N)$, where $g_J = (2J+1)$, $J$ is the rotational quantum number, $g_N = (2I+1)$ is the nuclear spin degeneracy, and $I$ is the nuclear spin (equal to 3
for states of odd $J$ and 1 for states of even $J$).}
Square symbols indicate the measured values, while the lines show the best two-component fit to the observations (see Table 3).  The $J = 8$ column density cannot be measured reliably in W44, W28, or 3C391, the H$_2$ S(6) line being blended with a strong 6.2~$\mu$m PAH emission feature.}
\end{figure}

\clearpage

\begin{figure}
\includegraphics[scale=0.95,angle=0]{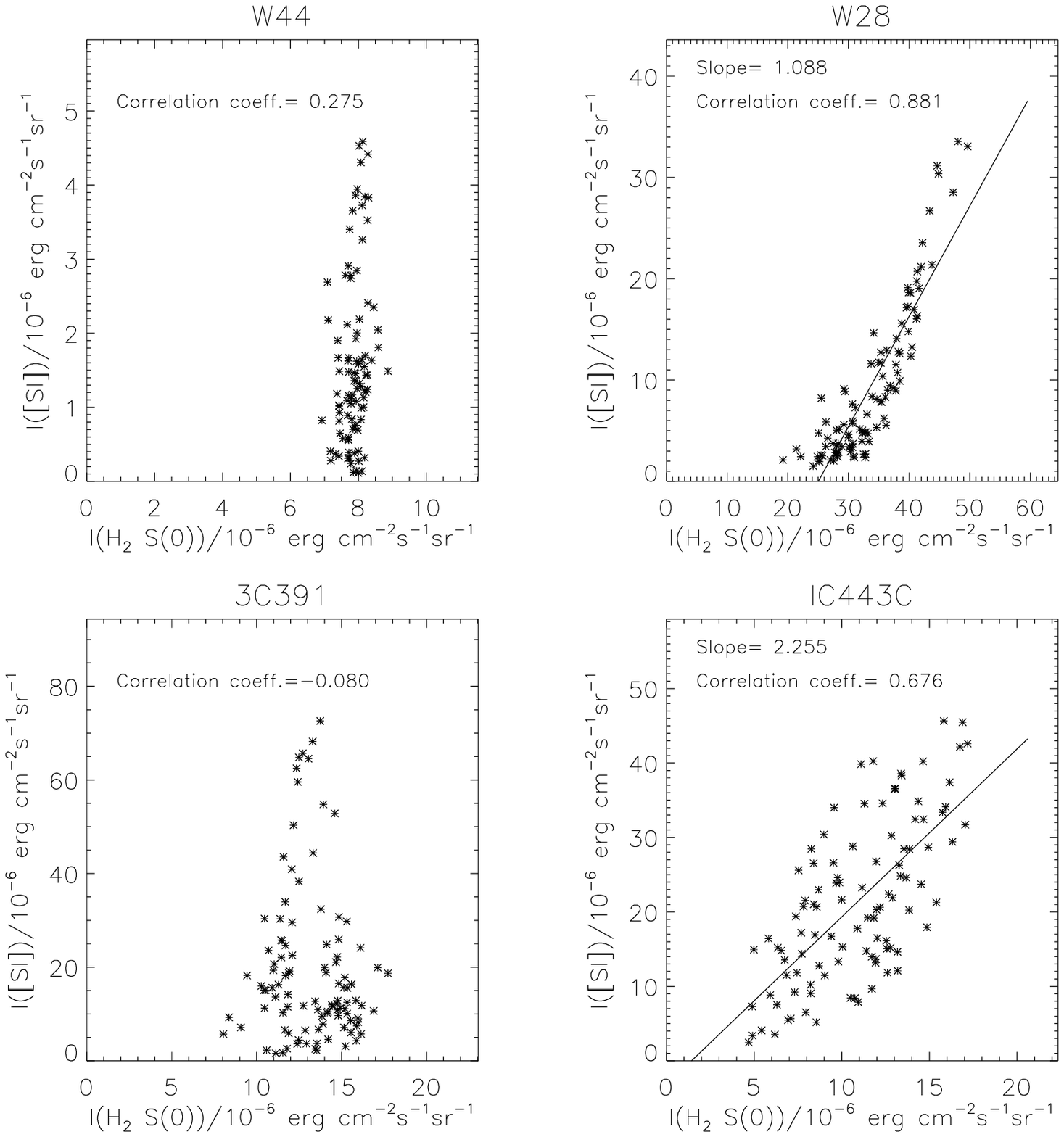}
\noindent{Fig.\ 19 -- Scatter diagrams showing correlations between [SI] and H$_2$ S(0).  Each point shows the mean intensities within a $5^{\prime \prime} \times 5^{\prime \prime}$ square region. }
\end{figure}
\clearpage

\begin{figure}
\includegraphics[scale=0.95,angle=0]{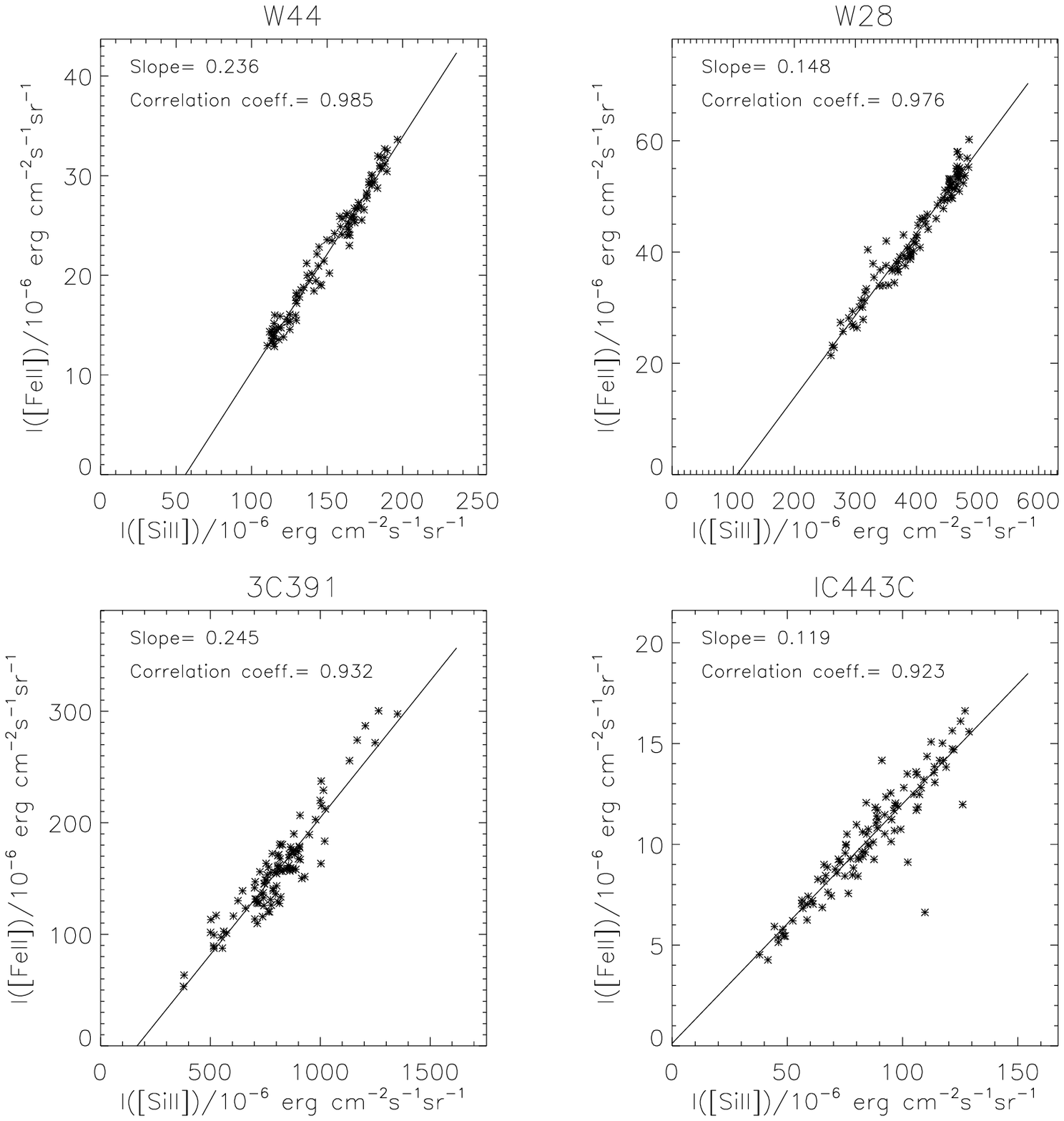}
\noindent{Fig.\ 20 -- Scatter diagrams showing correlations between the [FeII] 26 $\mu$m and [SiII] fine structure lines.  Each point shows the mean intensities within a $5^{\prime \prime} \times 5^{\prime \prime}$ square region. }
\end{figure}
\clearpage

\begin{figure}
\includegraphics[scale=0.95,angle=0]{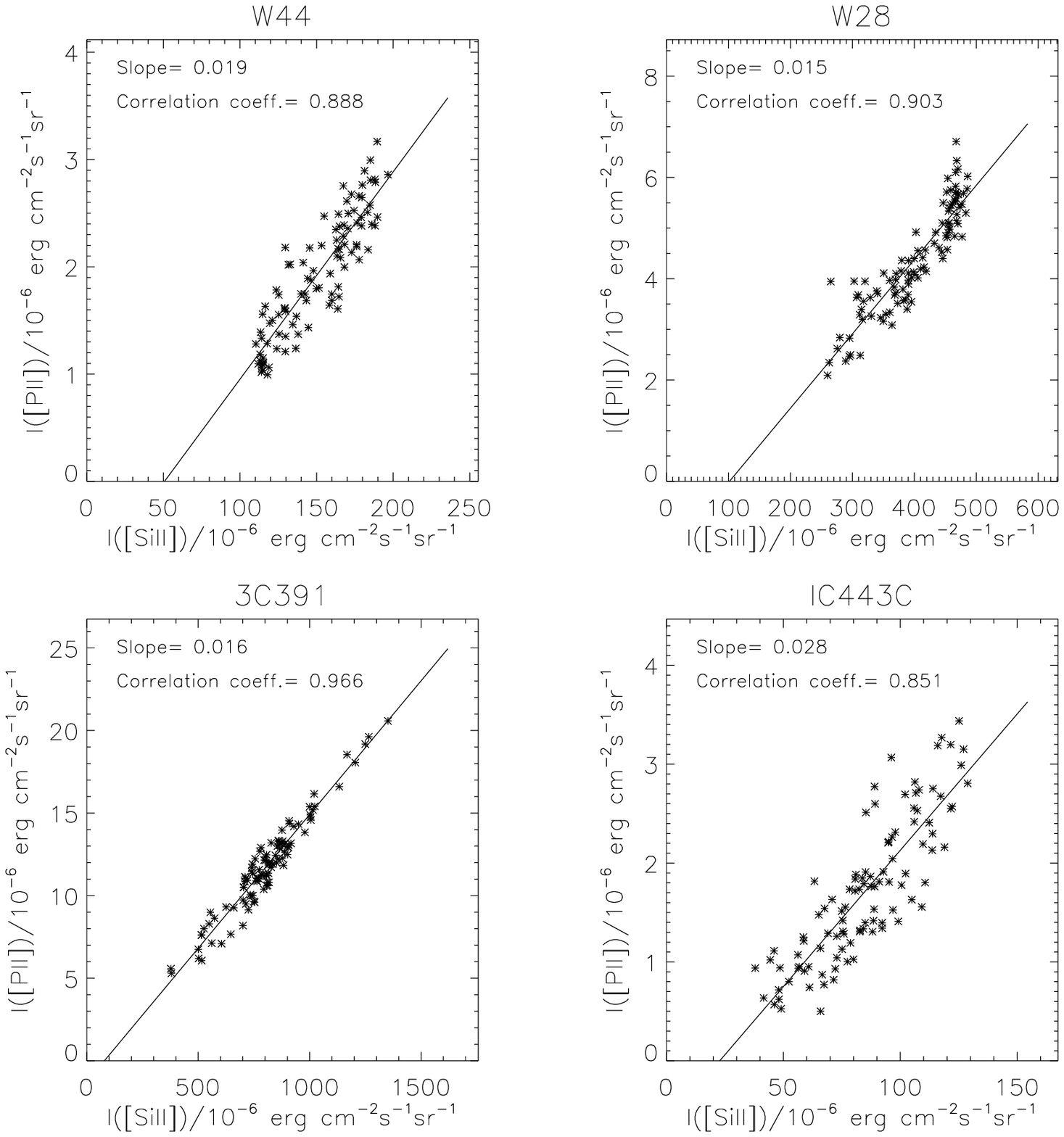}
\noindent{Fig.\ 21 -- Scatter diagrams showing correlations between the [PII] 26 $\mu$m and [SiII] fine structure lines.  Each point shows the mean intensities within a $5^{\prime \prime} \times 5^{\prime \prime}$ square region. }
\end{figure}
\clearpage

\end{document}